\documentclass{aa}  

\usepackage{bm}
\usepackage{subfig}
\usepackage{rotating}
\usepackage{amssymb,amsmath}
\usepackage{graphicx}
\usepackage{txfonts}
\usepackage[normalem]{ulem}

\begin{document} 

   \title{RSM detection map for direct exoplanet detection in ADI sequences}

   \author{C.-H. Dahlqvist\inst{1} \and F. Cantalloube\inst{2} \and O. Absil\inst{1}}

   \institute{STAR Institute, Universit\'{e} de Li\`{e}ge, All\'{e}e du Six Ao\^{u}t 19c, 4000 Li\`{e}ge, Belgium\\
              \email{carl-henrik.dahlqvist@uliege.be} \and
              Max Planck Institute for Astronomy, Königstuhl 17, 69117 Heidelberg, Germany \vspace{5pt}\\
                            Received ; accepted}
              
   \date{}

\abstract
   {Beyond the choice of wavefront control systems or coronographs, advanced data processing methods play a crucial role in disentangling potential planetary signals from bright quasi-static speckles. Among these methods, angular differential imaging (ADI) for data sets obtained in pupil tracking mode (ADI sequences) is one of the foremost research avenues, considering the many observing programs performed with ADI-based techniques and the associated discoveries.}
   {Inspired by the field of econometrics, here we propose a new detection algorithm for ADI sequences, deriving from the regime-switching model first proposed in the 1980s.}
   {The proposed model is very versatile as it allows the use of PSF-subtracted data sets (residual cubes) provided by various ADI-based techniques, separately or together, to provide a single detection map. The temporal structure of the residual cubes is used for the detection as the model is fed with a concatenated series of pixel-wise time sequences. The algorithm provides a detection probability map by considering two possible regimes for concentric annuli, the first one accounting for the residual noise and the second one for the planetary signal in addition to the residual noise. }
   {The algorithm performance is tested on data sets from two instruments, VLT/NACO and VLT/SPHERE. The results show an overall better performance in the receiver operating characteristic space when compared with standard signal-to-noise-ratio maps for several state-of-the-art ADI-based post-processing algorithms.}
   {}


   \maketitle
%

\section{Introduction}
\label{sec:intro}

High contrast imaging (HCI) is one of the most challenging techniques for exoplanet detection, but is also one of the most promising \citep[see][for a review]{Bowler16}.
The main difficulties encountered with HCI arise from the small angular separation between the host star and the potential exoplanets, the flux ratio between them (usually below $10^{-3}$), and the image degradation caused by the Earth's atmosphere. 
Adaptive optics (AO) and coronagraphic techniques are now widely used to improve the quality and reduce the dynamic range of the images in dedicated instruments such as GPI \citep{Macintosh08}, SPHERE \citep{Beuzit19}, or SCExAO \citep{Lozi18}. However, despite the use of these cutting-edge technologies, the resulting images are 
still affected by residual aberrations.  Under good observing conditions, the performance of HCI instruments is limited by aberrations arising in the optical train of the telescope and instrument, generating quasi-static speckles in the field of view.
Different processing techniques along with observing strategies have been proposed in the last decade to deal with these quasi-static speckles, whose shape and intensity (about $10^{-4}$) are similar to potential companions.

Angular differential imaging \citep[ADI,][]{Marois06} is nowadays the most commonly used observing strategy to mitigate quasi-static speckles in HCI. This observing strategy consists in acquiring images in pupil tracking mode, that is, with the instrument derotator keeping the pupil orientation fixed. The aim of this approach is to keep the quasi-static speckles fixed in the focal plane, so that they can easily be identified with respect to astrophysical objects rotating around the star along with the parallactic angle. Using this temporal diversity, a model of the speckle field, often referred to as the reference point spread function (reference PSF), may be built from the data. This reference PSF is then subtracted from the set of ADI images. The resulting residuals frames are eventually aligned and combined to detect the signal of potential exoplanets or discs, which should not have suffered too much from subtraction of the reference PSF.

Several methods using this approach have been proposed to maximise the noise reduction, such as for example a locally optimised combination of images \citep[LOCI,][]{Lafreniere07}, principal component analysis \citep[PCA,][]{Soummer12}, non-negative matrix factorization \citep[NMF,][]{Ren18}, and low rank plus sparse decomposition \citep[LLSG,][]{Gonzalez16}, allowing the user to reach contrasts down to $10^{-6}$ at 0.5 arcsec in the H-band (1.6 microns) with the latest generation of HCI instruments \citep[e.g.][]{Vigan15}. Another family of post-processing algorithms replaces the reference PSF subtraction by a forward modelling of the planetary companion using an inverse problem framework (ANDROMEDA, \citealt{Cantalloube15}; FMMF, \citealt{Pueyo16}, \citealt{Ruffio17}). In both cases, the detection is typically performed via the estimation of signal-to-noise-ratio (S/N) maps. In contrast with the forward- model-based algorithms, which provide a S/N map as a by-product of the model, in the case of reference PSF subtraction the S/N map is usually generated by using the median-averaged residual frames to estimate the annulus-wise S/N of every pixel it contains. The detection of planetary candidates is then done via the definition of an S/N threshold. Several methods have been proposed to generate S/N maps from the set of reference PSF-subtracted images \citep[e.g.][]{Mawet14,Bottom17,Pairet19}.

In this paper, we propose a novel approach to dealing with this last step of the ADI sequence post-processing. Instead of averaging the set of de-rotated images obtained after the reference PSF subtraction and computing an S/N map, we propose to consider the entire set of residual frames and rely on a regime-switching algorithm to classify the pixel values into two categories, regrouping either the planetary signals or the quasi-static speckles. The probability associated with the planetary regime then allows the creation of a detection map. The algorithm derives from the Markov regime-switching model first proposed by \citet{Hamilton88}, which is widely applied to analyse economic and financial time series. The aim of our new detection algorithm is to more effectively treat the residual noise still observed in the cube of residuals provided by ADI methods, increasing our ability to disentangle faint signals from bright speckles. The flexibility of the algorithm allows the use of ADI cubes treated with most post-processing methods. The cubes of residuals obtained from the different post-processing methods may be used separately, but can also be used  together,  further improving the sensitivity of the detection algorithm to faint companions.

The rest of the paper is organised as follows. In Section 2, we describe the new regime-switching model for the detection of exoplanets. Section 3 presents in detail the model estimation and the definition of the different parameters. The ability of our model to disentangle faint planetary signals from bright speckles is tested in Section 4 by injecting fake companions into two different data sets and by comparing the results with state-of-the-art ADI-based post-processing techniques. Finally, Section 5 concludes on this work.

\section{Regime-switching model}
\label{sec:model}

 The proposed detection algorithm derives from the Markov-switching regressions introduced by \citet{Goldfeld73} and \citet{Cosslett85} and further improved by \citet{Hamilton88,Hamilton94}, who developed an iterative inference algorithm to estimate the model parameters, namely the Markov regime-switching model (RSM). This approach is one of the most popular non-linear time series models in the econometric literature and many variants have been proposed. The aim of the RSM is to take into account possible dramatic changes in the behaviour of time series such as the transition between economic expansion and contraction in the case of financial time series. The regime-switching model relies on several linear equations to describe the different states of a system described by a time series. The probability of being in a given state depends on both a pre-defined transition probability and on the ability of the different equations to properly describe the evolution of the time series. One of the model outcomes is the probability associated with different regimes.  For each element of the time series, the
RSM provides the probability of being in any one of the different regimes. Our detection map derives directly from these probabilities.

In the case of our RSM detection map, the time series is built from the de-rotated cube of residuals obtained after the PSF subtraction and de-rotation steps of the ADI sequence post-processing. Several cubes of residuals treated with different ADI PSF subtraction techniques may be stacked in the time axis to provide additional information and increase the ability of the model to detect faint companions. To allow for the detection of planetary signals, we rely on two different regimes to model the de-rotated cube of residuals: a regime in which the residuals time series is described by speckle noise and a second regime with speckle noise plus a planetary signal. The planetary signal may be modelled as the measured off-axis PSF\footnote{For coronagraphic imaging, an off-axis non-coronagraphic image of the target is routinely acquired before and after the observing sequence. This PSF reference is used to calibrate the flux of the star and provide a model of the planetary signal for a forward model-based algorithm. For non-coronagraphic imaging, this reference PSF is the unsaturated exposure.} or as a forward model of the off-axis PSF after the subtraction. We consider in this paper the measured off-axis PSF for simplicity, although the algorithm may be easily adapted to a forward modelled off-axis PSF. 

The RSM we propose here is a modified version of the original Markov-switching model, in which only one parameter is determined via a maximum log-likelihood estimation. We rely on the characteristics of the data set to define the other model parameters. Having presented the basic principles behind our RSM, we may now describe the detailed procedure for our RSM detection map computation.

\begin{figure}[t!]
\center

\includegraphics[width=250pt]{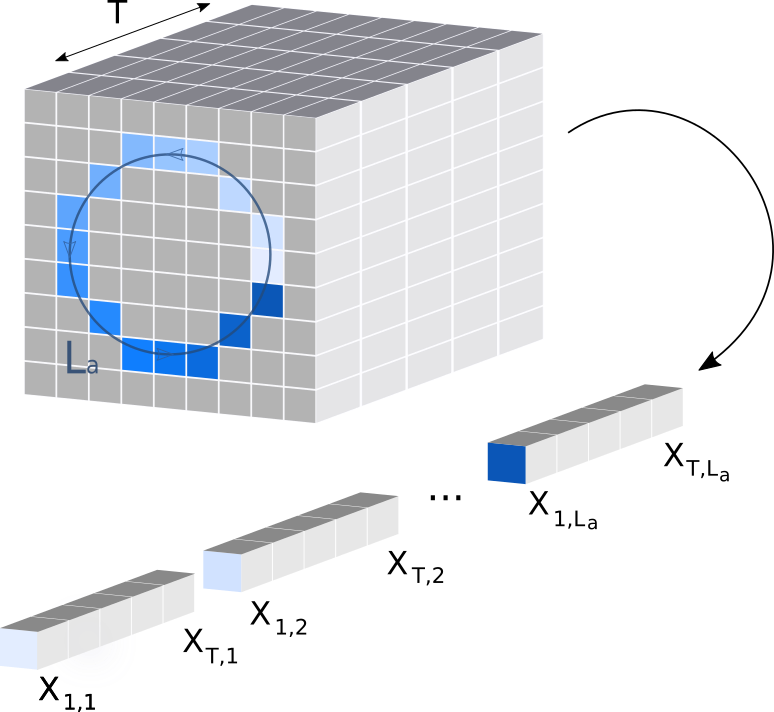}

                        \caption{Residuals time series for a given annulus $a$ is obtained by stacking the pixel values of the considered annulus along the time axis.}
                        \label{stacking}
                                \end{figure}

\subsection{Building the time series}

The first step of our estimation procedure is to build the time series that the regime switching model will try to model. As the noise properties are expected to evolve with radial distance, the regime switching model is applied annulus-wise.  For each annulus $a$, a specific residuals time series $X_{i_a}$ is built by vectorizing that part of the cube of residuals, indexed by $i_a$ the flattened pixel number. The length of the time series $X_{i_a}$ depends on the number of pixels in the considered annulus $L_a$  but also on the number of frames in the original de-rotated cube of residuals $T$. We indeed take advantage of all the individual frames contained in the de-rotated cube of residuals instead of collapsing the cube as is usually done when estimating an S/N map. As can be seen from Fig.~\ref{stacking}, the time series $X_{i_a}$ is built by concatenating the set of $T$ observations for every pixel contained in the annulus $a$, i.e. $X_{i_a}=\{X_{1,1},X_{2,1},\dots, X_{T,1},X_{1,2},\dots, X_{T,2},\dots,X_{T,L_a} \}$ with $i_a \in \{1,\dots, T \times L_a \}$. The first subscript of $X$ indicates the selected frame in the de-rotated cube of residuals, while the second one provides the position of the considered pixel in the selected annulus $a$. Both subscripts are replaced by a single index $i_a$ to form the residuals time series that feeds the RSM.

We consider first the time axis and then the spatial axis in order to stay  in the planetary regime during $T$ steps of the iterative process used to build the detection map, instead of switching $T$ times between both regimes when a planetary signal is present in a given annulus. Indeed, when travelling through the residuals time series, the planetary signal observed in a given pixel will act on the regime-switching model during $T$ steps, allowing the probability of being in the planetary regime to build up thanks to the short-term memory of the model. This helps to enhance the sensitivity of the algorithm to faint signals as it allows the probability to build up for a longer period of time.

\begin{figure}[t!]
\center

\includegraphics[width=150pt]{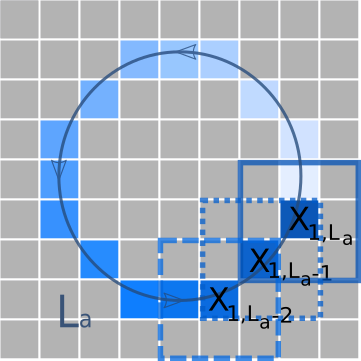}

                        \caption{Residuals matrices obtained from the first frame of the cube of residuals for the last three pixels of the annulus with $\theta$ equal to $3$. The time series $\bm{X}_{i_a}$ is created
by considering matrices of dimension $\theta \times \theta$ centred on every
$X_{i_a}$  in the cube of residuals.}
                        \label{timeseries}
                                \end{figure}

\subsection{Model description}

The second step of the RSM detection map computation consists in defining the set of equations describing the residuals time series for the two considered regimes. In the first regime, the time series $\bm{X}_{i_a}$ is described by a residual noise following the statistics of the quasi-static speckle residuals contained in the annulus. In the second regime, the time series $X_{i_a}$ is described by both the residual noise and the planetary signal model (off-axis PSF). The PSF being two-dimensional, we consider not only one pixel at a time but a batch of pixels in a square of size $\theta$ equal to the full width at half maximum (FWHM) of the PSF. In order to define the probability of observing a planetary signal at a given pixel $X_{i_a}$, we therefore need to consider a number of neighbouring pixels depending on the value of $\theta$. As depicted in Fig.~\ref{timeseries}, we define $\bm{X}_{i_a}$ as the the residuals matrices of dimension $\theta \times \theta$ centred on $X_{i_a}$, which will replace the time series $X_{i_a}$ used so far. Larger values of $\theta$ may be considered in the case of a forward-modelled off-axis PSF to take into account the signal self-subtraction, which, for instance, could create negative wings in the azimuthal direction. Our RSM is therefore characterised by the following equations:

\begin{eqnarray}
\label{maineq}
\bm{X}_{i_a} = \mu+ \beta R_{i_a} \bm{P}+ \bm{\varepsilon_{s,i_a}} =
  \begin{cases}
    \mu+  \bm{\varepsilon_{0,i_a}}  & \quad \text{if } S_{i_a}=0\\
    \mu+ \beta \bm{P}+ \bm{\varepsilon_{1,i_a}}      & \quad \text{if } S_{i_a}=1,\\
  \end{cases}
\end{eqnarray}
where $\beta$ provides the strength of the planetary signal, $\mu$ the mean of the quasi-static speckle residuals, and $\bm{\varepsilon_{s,i_a}}$ their time and space varying part characterised by the quasi-static speckle residuals statistics (see Table \ref{variable} for a summary of all the variables used in the RSM). Here, $ \bm{P}$ is the model of the planetary signal, which is the normalised off-axis PSF in the FWHM region.

As can be seen from Eq.~\ref{maineq}, there exist two possible states $S_{i_a}$, which are reflected in the value taken by the parameter $R_{i_a}$, with $R_{i_a} =1$ in the case of a planetary signal detection and $R_{i_a} =0$ in the other case. Here, $S_{i_a}$ is not directly observable, but we see its effect on the behaviour of $\bm{X}_{i_a}$ via the realisation $R_{i_a}$. 

The parameter $R_{i_a}$ is a realisation of a two-state Markov chain allowing short-term memory. This implies that we only consider the state $S_{i_a-1}$ in which the system was at index $i_a-1$ to define the probability of being in a given state $S_{i_a}$ for the current index $i_a$. The fact that the realisation $R_{i_a}$ is a probabilistic outcome implies that we cannot consider being in only one of the two regimes. We have instead a given probability of being in each of them. Our RSM tries to describe the behaviour of the time series $\bm{X}_{i_a}$ via a probability-weighted sum of the values generated by the equation describing each regime.

\subsection{Definition of the model probabilities}

The probability of $\bm{X}_{i_a}$ being in a state or regime $S_{i_a}=s$ is characterised by the set of parameters of Eq.~\ref{maineq}, that is, $P$ the planetary signal model, and $\mu$ and $\beta$, the statistical properties of the residual noise $\bm{\varepsilon_{s,i_a}}$. We make the simplifying assumption here that the quasi-static speckles residuals $\bm{\varepsilon_{s,i_a}}$ may be characterised to a good level of precision by their mean $\mu$ and variance $\sigma$. We write the probability of observing $\bm{X}_{i_a}$ in the state $s$ at step $i_a$ as follows: 

 \begin{eqnarray}
\xi_{s,i_a}= \mathrm{P}(S_{i_a}=s | \Omega_{i_a},\bm{P}, \mu,\beta,\sigma),
\end{eqnarray}
where $ \bm{P}$,  $\mu$, $\beta$, $\sigma$ and $\Omega_{i_a}=\{\bm{X}_{i_a},\bm{X}_{i_a-1}\}$ provide the parameters of the model.

This probability $\xi_{s,i_a}$ is the key element of our RSM detection map as the map is constructed based on the value taken by $\xi_{1,i_a}$ for every pixel of every annulus. Indeed, $\xi_{1,i_a}$ provides a detection probability for each pixel and each frame of the de-rotated cube of residuals. The final RSM detection map is created by averaging these probabilities along the time axis of the cube of residuals. 

In the case of a two-state Markov chain, the computation of  $\xi_{s,i_a}$ necessitates the estimation of (i) the probability $\xi_{q,i_a-1}$ of observing the system in the state $q$ at step $i_a-1$, (ii) the transition probability $p_{q,s}$ from state $q$ to state $p$ and (iii) the likelihood of observing $\bm{X}_{i_a}$ in state $s$ at step $i_a$, which we note $\eta_{s,i_a}$. The probability of being in a state $s$ at index $i_a$ can be computed as the normalised likelihood of being in state $s$ at index $i_a$ multiplied by the probability of having been in either of the two states at index $i_a-1$ and by the transition probability $p_{q,s}$, which accounts for the short-term memory of the algorithm. The expression of the state probability $\xi_{s,i_a}$ is therefore given by the following expression \citep{Hamilton88}:

\begin{eqnarray}
        \label{estxsi}
\xi_{s,i_a}=\sum^{1}_{q=0} \frac{\eta_{s,i_a} p_{q,s} \; \xi_{q,i_a-1} }{f(\bm{X}_{i_a}| \Omega_{i_a-1}, \bm{P}, \mu,\beta,\sigma)} ,
\end{eqnarray}
with the sum $f$ of conditional densities for index $i_a$ given by:

\begin{eqnarray}
f(\bm{X}_{i_a}| \Omega_{i_a-1},\bm{P}, \mu,\beta,\sigma)= \sum^{1}_{q=0} \sum^1_{s=0} \eta_{s,i_a} p_{q,s} \; \xi_{q,i_a-1}, 
\end{eqnarray}
and the transition probabilities given by:

\begin{eqnarray}
p_{q,s}= \mathrm{P}(S_{i_a}=s \mid S_{i_a-1}=q) ,
\end{eqnarray}
with $q,s \in \{0,1 \} $. We consider the two possible states describing the system at index $i_a-1$ via the sum over $q$. The function $f(\bm{X}_{i_a}| \Omega_{i_a-1},\bm{P}, \mu,\beta,\sigma)$, which represents the numerator summed over the two possible states taken at index $i_a$, ensures that the sum of the probability $\xi_{s,i_a}$ equals one for every index $i_a$. 

\subsection{Transition probabilities estimation}
\label{subsec:transprob}

For our two-regime model, the transition probability $p_{q,s}$ regroups the probabilities of staying in either regime along with the probabilities of switching to the other regime. The estimation of $p_{q,s}$ is relatively straightforward by imposing on the algorithm the potential existence of no more than one planetary signal per annulus. A number of planetary signals per annulus in the interval $\left] 0,1\right]$ may therefore be considered. Following our testing, a value of one companion per annulus must be privileged in the case of faint companions as lower values decrease both the residual speckles and the companion intensities in our model. Considering the number of pixels $L_a$ and the number of frames $T$, the parametrisation of $p_{q,s}$ translates as follows in the case of one planetary signal per annulus:

\begin{eqnarray}
p_{q,s} =
 \begin{pmatrix}
 p_{0,0}= 1-1/(T \times L_a) & p_{1,0}= 1/T   \\
 p_{0,1}=1/(T \times L_a)   & p_{1,1}= 1-1/T  \\
 \end{pmatrix}.
\end{eqnarray} 

\subsection{Likelihood function definition}

The determination of the likelihood is the key step of the model estimation. The challenge is to select the right probability distribution function to  properly describe $\bm{\varepsilon_{s,i_a}}$, the residual noise due to the quasi-static speckles. Indeed, the value taken by $\eta_{s,i_a}$ depends directly on the position of the elements of $\bm{X}_{i_a}$, or the elements of $\bm{X}_{i_a} -\beta \bm{P}$, in the probability distribution of the quasi-static speckle residuals. Considering the small transition probabilities $p_{0,1}$, the probability of planetary signal detection $\xi_{1,i_a}$ depends heavily on the value taken by $\eta_{1,i_a}$. The parametrisation of the selected probability distribution function also plays  an important role. 

Different probability distribution functions may be used. For the sake of clarity, we illustrate the likelihood function definition with a simple Gaussian distribution as is done in \citet{Hamilton88}. However, the following section will allow us to investigate the question of the optimal probability distribution function selection as different post-processing algorithms provide different noise distributions for different separations. The Gaussian distribution allows us to construct a likelihood function for state $s$ at index $i_a$ in the following manner:

 \begin{eqnarray}
\eta_{r,i_a}= \sum^{\theta^2}_n \frac{1}{\theta^2} \frac{1}{\sqrt{2 \pi}\sigma} \exp\left[- \frac{(\bm{X}^n_{i_a} - R_{i_a} \beta \bm{P}^{n} -\mu)^2}{2\sigma^2}\right],
\end{eqnarray}
with $n$ the index of the matrix elements for $\bm{X}_{i_a}$ and $\bm{P}$.\ The sum over the matrix elements allows us to obtain only one value per considered $\theta \times \theta$ patch.

\subsection{Model estimation}

Since the estimation of $\xi_{s,i_a}$ depends on its value at the previous step, we rely on an iterative procedure to estimate the entire set of $\xi_{s,i_a}$. This iterative procedure requires the definition of an initial condition for $\xi_{q,0}$. Assuming that the considered Markov chain is ergodic, we can simply set $\xi_{q,0}= \mathrm{P}(S_t=q \mid  \bm{P}, \mu,\beta,\sigma)$ equal to the unconditional probability $\xi_{q,0}= \mathrm{P}(S_t=q)$. Following the approach proposed by \citet{Hamilton94}, the two initial probabilities $\xi_{0,0}$ and $\xi_{1,0}$ may be estimated using the following system of equations:
\begin{eqnarray}
\label{eqsyst}
  \begin{cases}
    \xi_{0,0} = \xi_{0,0} p_{0,0} + \xi_{1,0}p_{1,0}\\
    \xi_{1,0} = \xi_{1,0} p_{1,1} + \xi_{0,0}p_{0,1}\\
    \xi_{1,0}+ \xi_{0,0}=1,\\
  \end{cases}
\end{eqnarray}
which translates in terms of matrices into:
\begin{eqnarray}
\bm{A} \bm{\xi}= \bm{\psi},
\end{eqnarray}
with $\bm{\epsilon}=\left[ \xi_{0,0},\xi_{1,0}\right]$ the set of initial probabilities,  $\bm{\psi}=\left[ 0,0,1\right]$ , and $\bm{A} $ given by:
 \begin{eqnarray}
\label{mat}
\bm{A}=
\begin{pmatrix}
 I_{2 \times 2} - P\\
1 \;\;\;\; 1
 \end{pmatrix},
\end{eqnarray}
with $P$ the matrix of $p_{q,s}$, $I_{2 \times 2}$ a diagonal matrix of dimension $2 \times 2$. Solving the system of equations (eq.\ref{eqsyst}) to obtain the initial probabilities, $\bm{\xi}$, is then equivalent to taking the third row of the matrix $(\bm{A}^t\bm{A})^{-1}\bm{A}^t$.

\begin{table*}[h]
                        \caption{Description of the mathematical notations for the variables used in the RSM detection map computation.}
                        \label{variable}
\centering

                        \begin{tabular}{lll}
                        \hline
                        \hline
Symbol  & Dimension & Comments\\                        
 \hline
$X_{i_a}$ &  $T L_a$&  Vector of residuals for the annulus $a$\\
$\bm{X}_{i_a}$ &$\theta \times \theta \times T L_a$ & Matrices of residuals centred on $X_{i_a}$\\
$a$& $1$ & Annulus index\\
$L_a$& $1$ & Number of pixels included in the annulus $a$\\
$T$& $1$ & Number of frames in the cube of residuals\\
$i_a$&$1$& Index associated with every pixel from every frame in the annulus $a$ (ranges from $1$ to $T L_a$)\\
$\theta$& $1$ & Angular size of the considered planetary signal ( set to $1$ $\lambda/D$)\\
$\mu$& $1$ & Mean of the residuals contained in an annulus $a$, with width equal to $\theta$\\
$\sigma$& $1$ & Standard deviation of the residuals contained in an annulus $a$, with width equal to $\theta$\\
$\beta$& $1$ &  Parameter representing the intensity of the planetary signal in the cube of residuals \\
$R_{i_a}$& $T L_a$ & Realisation of a two-state Markov chain representing the state in which the system is in for pixel $i_a$\\
$ \bm{P}$& $\theta \times \theta$ & Planetary signal (off-axis PSF)\\
$\bm{\varepsilon_{s,i_a}}$& $2 \times \theta \times \theta \times T L_a$ & Error terms associated with the two regimes\\
$S_{i_a}$& $T L_a$ & State in which the system is in for every pixel $i_a$\\
$\xi_{s,i_a}$& $2 \times T L_a$& Probability associated with state $s$ for every pixel $i_a$\\
$\eta_{s,i_a}$ & $2 \times T L_a$ & Likelihood of being in each state for every pixel $i_a$\\
$p_{q,s}$& $2 \times 2 $& Transition probabilities between the regimes\\
\hline
                        \end{tabular}
                                \end{table*}

\section{Detection map estimation}
\label{sec:estimation}

We propose in this section a procedure to produce a RSM detection map. The model we developed so far necessitates the computation of cubes of residuals along with the definition of several parameters:  the probability distribution function of the quasi-static speckles residuals $\bm{\varepsilon_{s,i_a}}$ and its first two moments, the planetary signal model $\bm{P}$, the intensity parameter $\beta$, and the transition probability $p_{q,r}$. The transition probability $p_{q,r}$ is already defined in Sect.~\ref{subsec:transprob}. We therefore consider the remaining three model parameters.

\subsection{Computation of de-rotated cubes of residuals} 
 
The first step to create a RSM detection map is the production of the de-rotated cubes of residuals for the selected ADI-based post-processing techniques feeding our regime-switching algorithm. As an illustration of the ability of our model to improve the detection when considering several methods at once,  in this paper we consider three different post-processing techniques: annular PCA, NMF, and LLSG.
For the two first approaches, the estimation of the cubes of residuals starts with the definition of a reference PSF. Annular PCA follows the PCA principles by computing the directions of maximal variance from the main matrix representing the ADI sequence, $\bm{M} \in \mathbb{R}^{n \times p} $, with $n$ the number of frames and $p$ the number of pixels in the considered annulus. The determination of a reference PSF is done via the estimation of the eigenvectors $\bm{V}$ of the matrix $\bm{M}$ by taking $\bm{V}_k$, the first $k$ components of $\bm{V}$. Annular PCA relies on  a separate estimation for each annulus composing the original cube of data to take into account the radial evolution of the noise distribution; it allows the user to consider the local structure of the speckle noise instead of the entire frame. The cube of residuals is then obtained via the subtraction of the low rank matrix $\bm{M}\bm{V}_k^T \bm{V}_k$ from the initial ADI sequence $\bm{M}$. 

As for annular PCA, NMF can be understood as a low rank approximation, with an additional non-negativity condition. This method consists in the decomposition of a matrix into two factors of non-negative values via the minimisation of the Frobenius norm:
 \begin{eqnarray}
\mathrm{argmin}_{\bm{W},\bm{H}} \frac{1}{2} \Vert \bm{M}- \bm{W}\bm{H} \Vert^2_{FN}= \frac{1}{2} \sum_{i,j} (M_{i,j}  -WH_{i,j} )^2,
\end{eqnarray}
where $\bm{W} \in \mathbb{R}^{n \times k} $ and $\bm{H} \in \mathbb{R}^{k \times p} $. The method allows the definition of a matrix $\bm{W}\bm{H}$ with rank $k$ lower than that of the original matrix $\bm{M}$, keeping only the main components of $\bm{M}$. The matrix $\bm{W}\bm{H}$ provides a reference PSF for the entire set of frames representing the structure of the residual starlight. As for annular PCA, this matrix is subtracted from the original ADI sequence to obtain the cube of residuals, $\bm{M}-\bm{W}\bm{H}$. 

Finally, the LLSG estimation is based on the decomposition of the cube intensities in three separate components: $\bm{L},$ a low-rank matrix, $\bm{S},$ a sparse matrix expected to contain the potential planetary signal, and $\bm{G},$ the Gaussian part of the background noise. This partly explains why the distribution of the resulting residuals observed in Fig.~\ref{resdistri} is far from being Gaussian, the Gaussian part of the noise having already been removed. More information about the algorithm may be found in \citet{Gonzalez16}. The cube of residuals is directly provided by $\bm{S}$.

\subsection{Probability distribution function}

\begin{figure*}[h!]
  \centering
  \subfloat[Annular PCA at $1\lambda/D$]{\includegraphics[width=220pt]{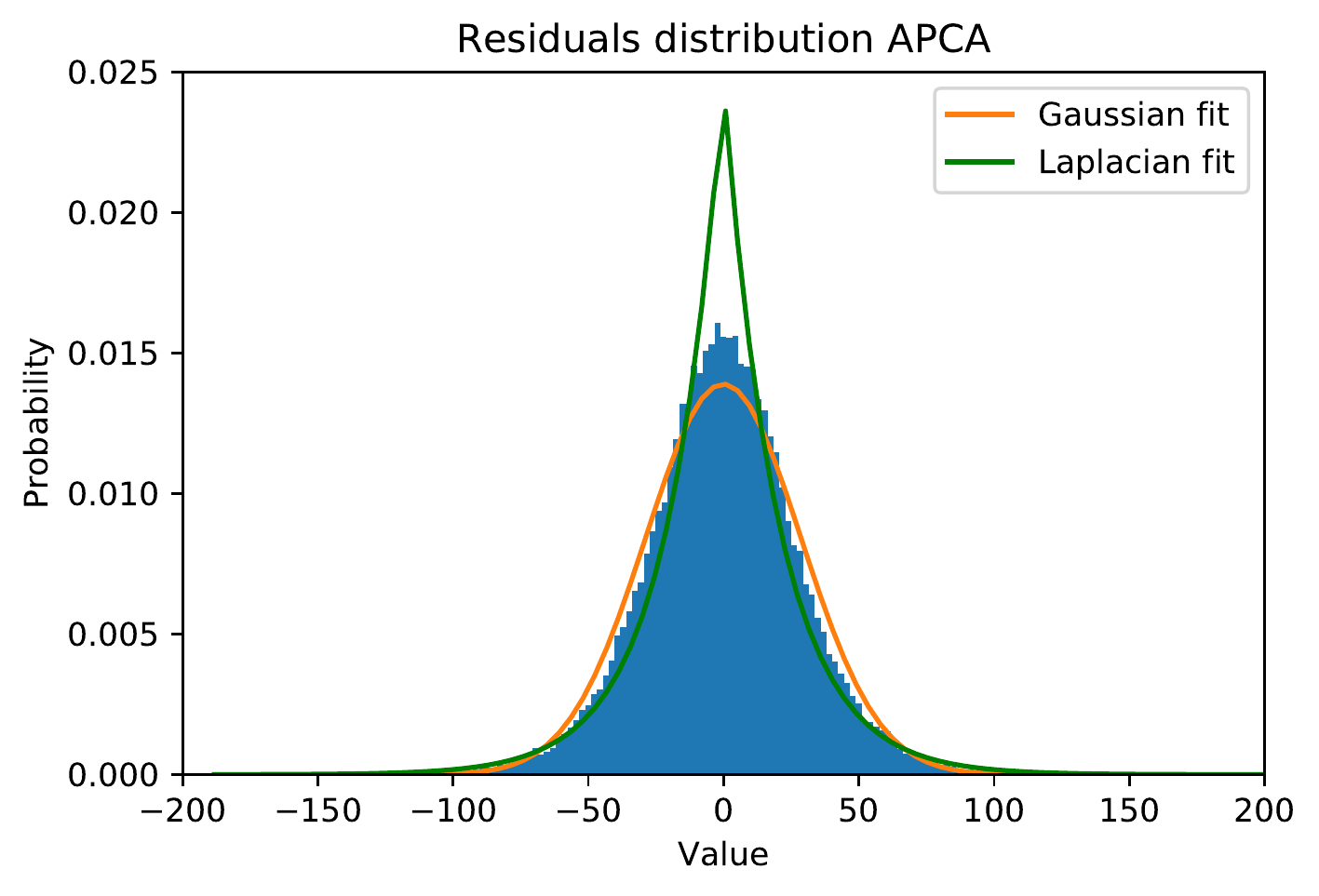}}
  \subfloat[ Annular PCA at $8$ $\lambda/D$]{\includegraphics[width=220pt]{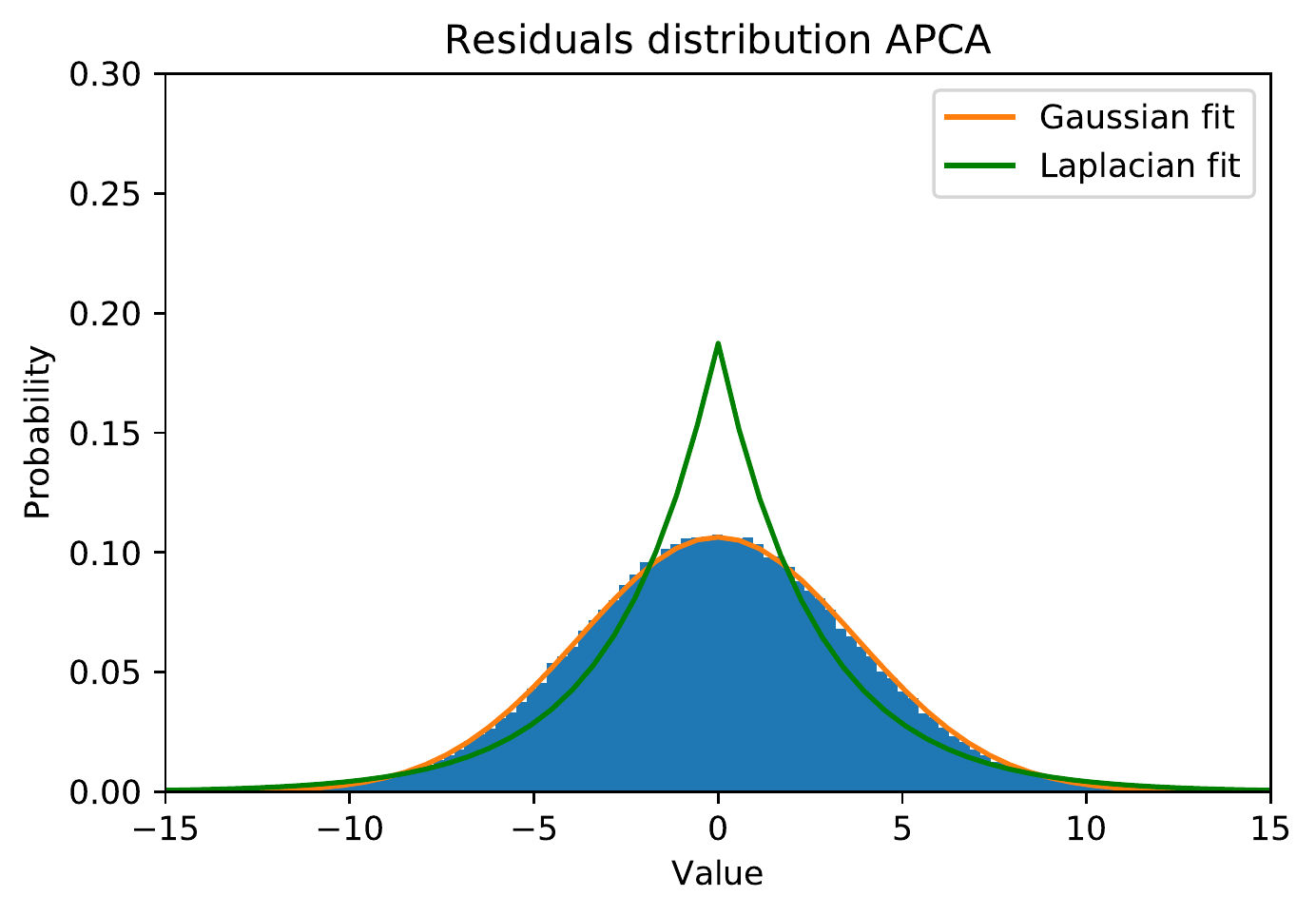}}\\
  \subfloat[NMF at $1\lambda/D$]{\includegraphics[width=220pt]{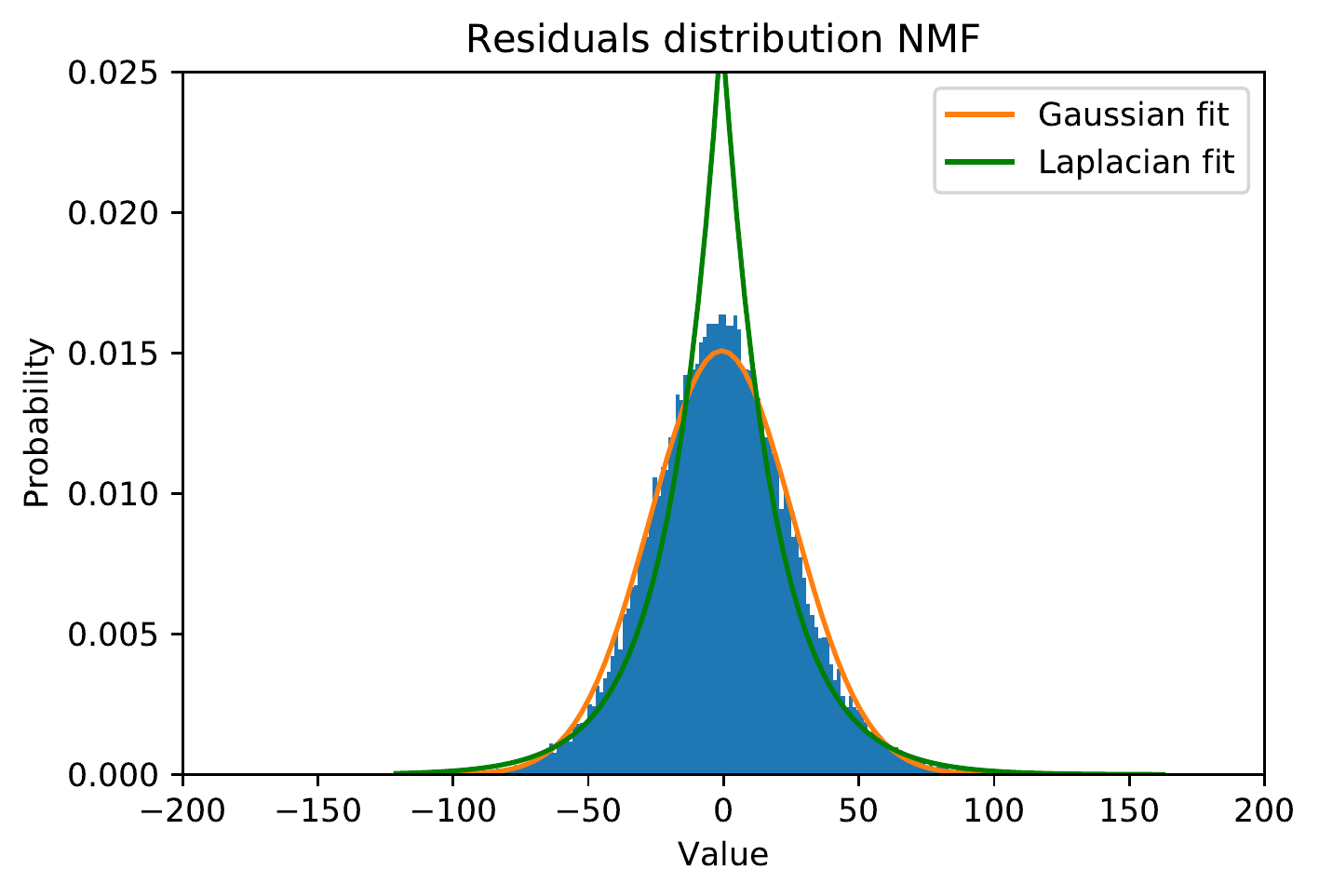}}
  \subfloat[NMF at $8$ $ \lambda/D$]{\includegraphics[width=220pt]{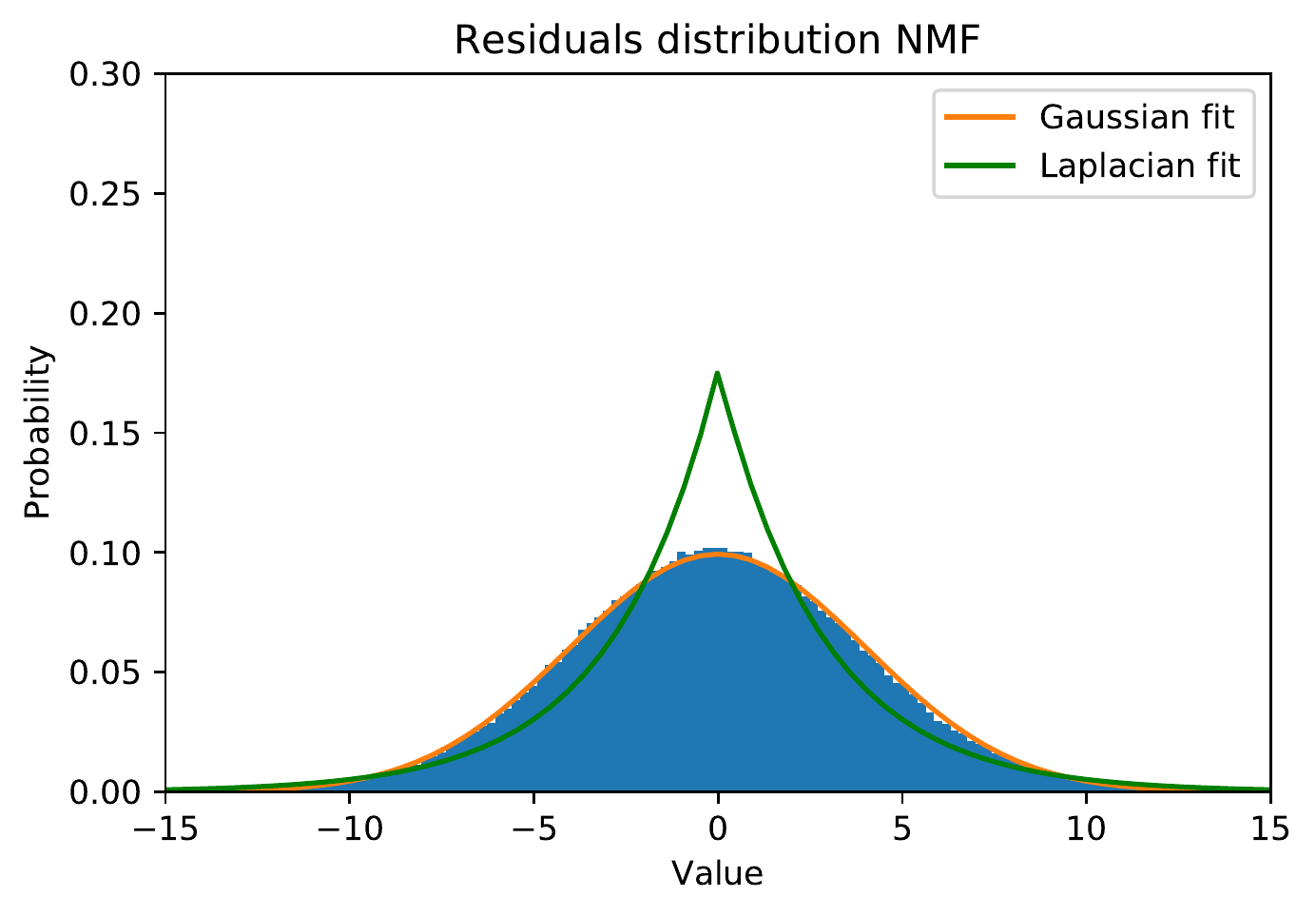}}\\
  \subfloat[LLSG at $1\lambda/D$]{\includegraphics[width=220pt]{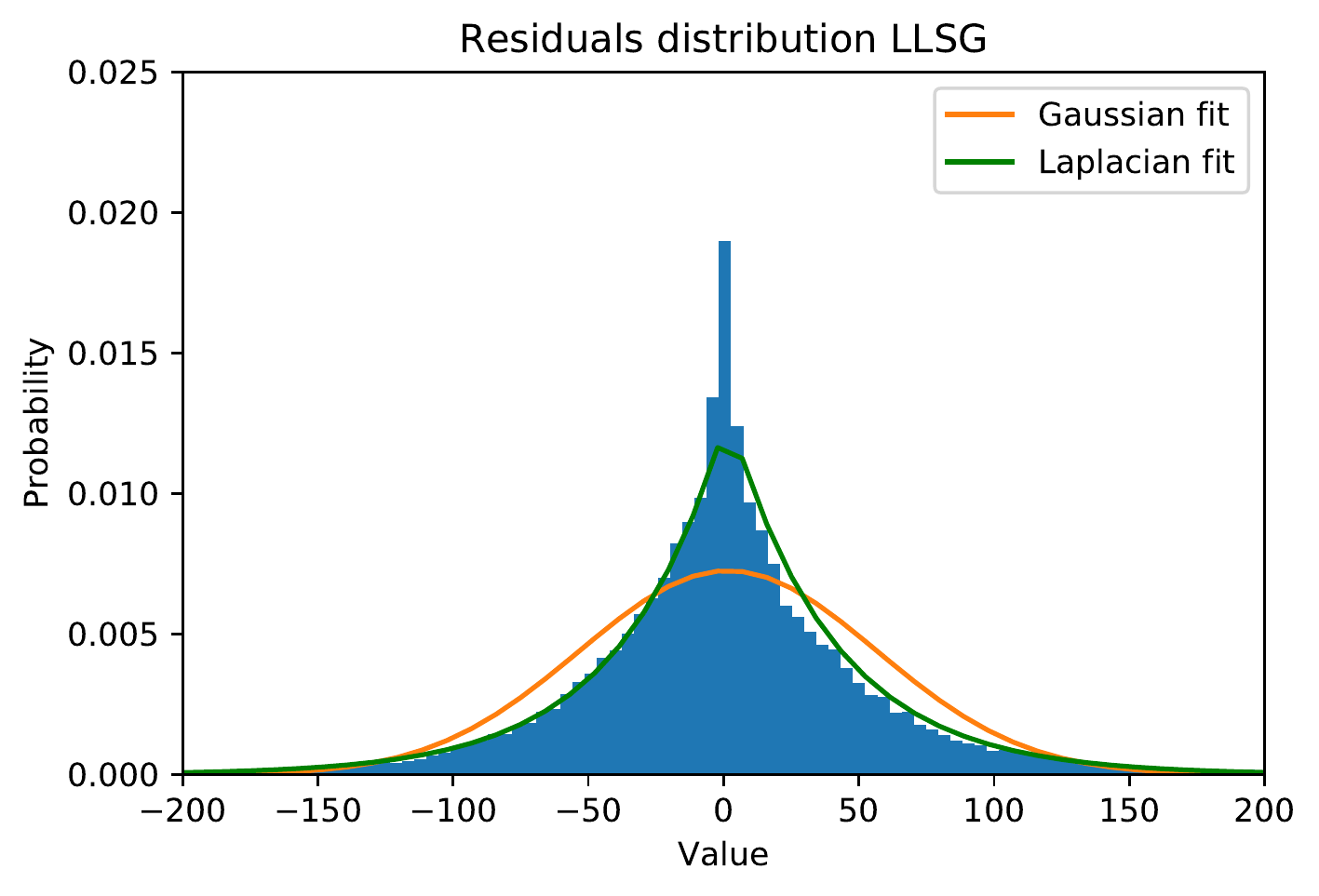}}
  \subfloat[LLSG at $8$ $ \lambda/D$]{\includegraphics[width=220pt]{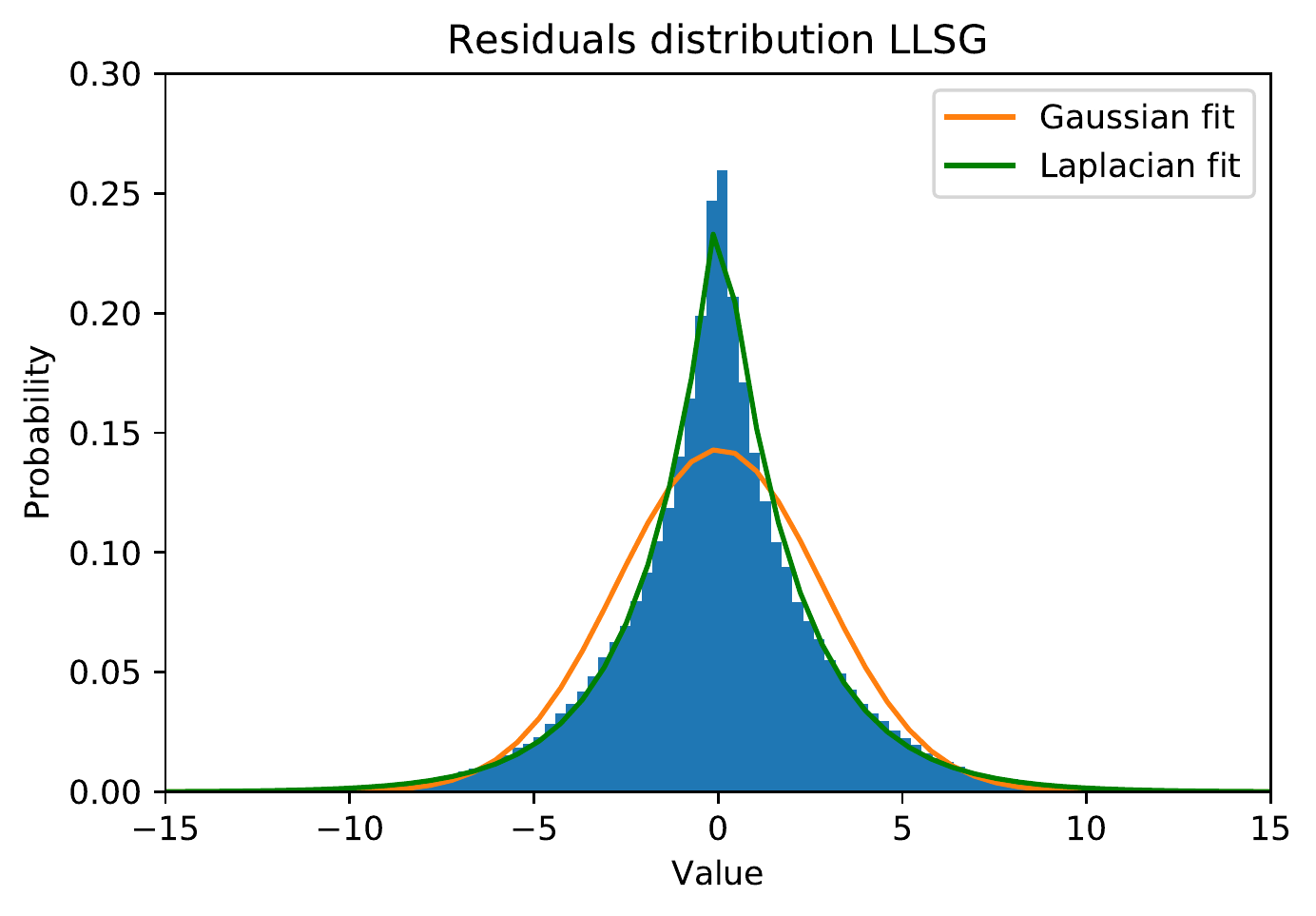}}\\ 
  
  \caption{\label{resdistri} Distribution of the residuals for a VLT/NACO data set after PSF subtraction by annular PCA (top), NMF (middle), and LLSG (bottom) along with a Gaussian (orange line) and Laplacian (green line) fit at small (left) and large separations (right), with respectively 20 components for the annular PCA and the NMF and a rank of 5 for the LLSG.}
\end{figure*}

We then move to the model parameters definition by first considering the selection of the probability distribution function describing the speckle residuals. Figure~\ref{resdistri} (a-d) provides the distribution of the residuals for a VLT/NACO ADI sequence (see Sect.~\ref{sec:simulation} for a description of the data set) obtained with respectively the annular PCA, the NMF, and the LLSG methods. We see from these graphs that the distribution of the residuals is either close to a Lapacian or to a Gaussian distribution depending on the selected post-processing techniques and on the angular separation. At small angular separations, the tails of the distributions of the residuals seem to be closer to a Laplacian, while at larger separation they seem closer to a Gaussian, except for LLSG processing. This radial evolution is mainly due to the higher (relative) number of intense speckles, the lower number of pixels, and the lower field rotation at small separation. Overall, the distribution of the residuals is close to a Gaussian for annular PCA and NMF, and close to a Laplacian for LLSG. This partially confirms the findings of \citet{Pairet19}, who demonstrated that the residuals were closer to a Laplacian than a Gaussian distribution, especially when looking at the tails of the distribution. 

The results of Fig.~\ref{resdistri} illustrate the difficulty of defining the residuals distribution as there exists a dependence on both the separation and the post-processing technique along with differences between the tails and the core of the distribution. We therefore consider both the Gaussian and Laplacian distributions in the performance assessment of Sect.~\ref{sec:simulation}.

The proposed regime-switching model provides a local detection probability as it considers one annulus at a time. The parameters of the residuals probability distribution should therefore be estimated locally. In the previous section, we considered not a single pixel at a time but a $\theta \times \theta$ matrix of pixels centred on the pixel of interest. We therefore estimate the pixel-wise mean and variance of the residuals empirically by considering an annulus with a width of $\theta$ pixels centred on the selected annulus. The entire set of frames is used for the estimation of these two parameters. Although planetary signal may be included in the annulus, the effect of this signal on the estimation of the mean and variance is limited and decreases with angular separation.

\subsection{Intensity parameter}

For the estimation of the intensity parameter $\beta$, we rely on the estimated variance of the pixel intensity in the annulus. We were inspired here by the S/N maps that are usually created with the final frame provided by most of the ADI techniques. We define the intensity parameter $\beta$ as a multiple of the estimated variance $\sigma$:
 \begin{eqnarray}
\beta = \delta \sigma.
\end{eqnarray}

 The $\beta$ parameter is the only parameter we propose to estimate via a maximum log likelihood. Several values of $\delta$ are tested in a given interval starting at $\delta=1$, as $\delta=0$ would imply a single regime model. The optimal $\delta$ in an annulus $a$ is the one leading to the highest log-likelihood sum $\sum_{i_a}^{L_a \times T} \log\left[f(\bm{X}_{i_a}| \Omega_{i_a-1},\bm{P}, \mu,\beta,\sigma)\right]$ in the considered interval.

Relying on this definition of $\beta$ allows us to obtain information about the position of the detected planetary signal inside the probability distribution of the residual speckles. A higher $\delta$ implies that the detected signal is farther in the distribution tails, which indicates a higher level of confidence (which will generally translate into a higher probability in the RSM map) about the detected planet and a higher flux for a given noise distribution. However, the $\beta$ parameter does not provide an estimation of the planetary flux as we are not using a forward model of the PSF for the planetary signal.

\subsection{Planetary signal model}

Using a forward model for the planetary signal would allow us to take into account the distortions (such as self-subtraction) created by ADI-based post-processing treatment when estimating cubes of residuals. Although a forward-modelled PSF should provide more accurate results, it should be noted that some ADI PSF subtraction techniques do not lend themselves to the analytical computation of a forward model (e.g. LLSG, NMF). A more universal numerical way to compute a forward model is to compare the initial cube of residuals and the one in which a fake companion has been injected. Following this approach, we tested numerical estimation of forward modelled PSF for generating RSM detection maps but without managing to improve the algorithm accuracy compared to the use of measured off-axis PSF. We therefore decided to only consider measured off-axis PSFs in the rest of this paper. However, a forward model variant of the proposed algorithm is still under development and should be a valuable improvement of the current model, at the expense of its computation time.

\subsection{Regime switching model detection map estimation} 

Now that we have defined the procedure to estimate the cubes of residuals feeding the RSM algorithm as well as the model parameters, we may summarise the main steps of the algorithm as follows: 

\begin{enumerate}

\item Compute the residuals cubes for the selected ADI techniques and de-rotate all the resulting frames;
\item define the separation to the star for the first and last annuli, respectively $a_{ini}=\mathrm{FWHM} /2 +1$ and $a_{fin}= (f_{size}-\mathrm{FWHM})/2$ with $f_{size}$ the size of the frame;
\item define the series $\bm{X}_{i_a}$ for the first annulus;
\item estimate the mean and variance of the residuals inside the annulus separately for each residuals cube;
\item using the iterative procedure described in Sect.~\ref{sec:model}, estimate $\xi_{r,i_a}$ for each index $i_a$ for the set of tested $\delta$;
\item include the probability of planetary signal $\xi_{1,i_a}$ providing the maximum likelihood in a three-dimensional matrix $\bm{U} \in \mathbb{R}^{L_a \times T}$; 
\item repeat steps 2 to 6 for the next annulus ($a+1$) until $a_{fin}$ is reached;
\item average the detection probability contained in $\bm{U}$ along the time axis to obtain the final RSM detection map.
\end{enumerate}

 The resulting detection map provides the averaged probability of observing a planetary signal in a given cube of data, along with the optimal $\beta$. The following section explores the effectiveness of this new approach when applied to observational data sets.

\section{Performance assessment}
\label{sec:simulation}

\subsection{Data}

We propose the use of two ADI sequences acquired with two instruments of the Very Large Telescope (VLT): NACO and SPHERE. This allows us to investigate the ability of our model to deal with the different noise profiles produced by these instruments.

 The first data set focuses on $\beta$ Pictoris and its planetary companion $\beta$ Pictoris b. It was obtained in $L'$ band in January 2013 with NACO in its AGPM coronagraphic mode \citep{Absil13}. The ADI sequence is composed of 612 individual frames obtained by averaging 40 successive individual exposures, each frame providing an effective integration time of 8 s. The parallactic angle ranges from -15$^{\circ}$ to +68$^{\circ}$. We use every third frame to reduce the CPU time and cropped  the central $101\times101$ pixels region to consider mainly the first arc-second. 

 The second data set is an ADI sequence on 51 Eridani produced by the SPHERE-IRDIS instrument using an apodized pupil Lyot coronagraph \citep{Samland17}. The sequence was taken in $K1$ band in September 2015 and regroups 194 frames with 16 s of integration time. The parallactic angle ranges from 297$^{\circ}$ to 339$^{\circ}$. The data set was pre-processed using the SPHERE Data Center pipeline \citep[for more details about the reduction see][]{Delorme17,Maire19}.

\subsection{Detection maps}

We start our analysis by considering the RSM detection map generated with the proposed algorithm and based on the residual cubes provided by annular PCA, NMF, and LLSG, and compare it with the S/N map obtained with the same three post-processing algorithms. The post-processing as well as the S/N detection maps are generated for all three methods with the \emph{VIP} package developed by \citet{Gonzalez17} using the standard parametrisation. Both annular PCA and LLSG are performed annulus-wise, with each annulus being divided into four segments in the case of LLSG. Other parametrisations are possible as the proposed approach works with any de-rotated cube of residuals. The three cubes of residuals obtained with the selected post-processing techniques are then stacked to create a single cube to feed the RSM. The variance and the mean of the residuals are estimated separately for each subcube as their noise profiles are specific, as demonstrated in the previous section when looking at the residual distributions.

\begin{figure*}
  \centering
  \subfloat[RSM Probability map]{\includegraphics[width=220pt]{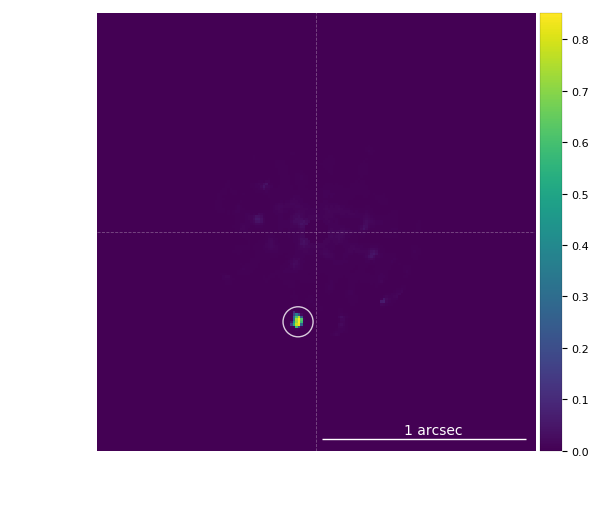}}
  \subfloat[Annular PCA S/N map]{\includegraphics[width=220pt]{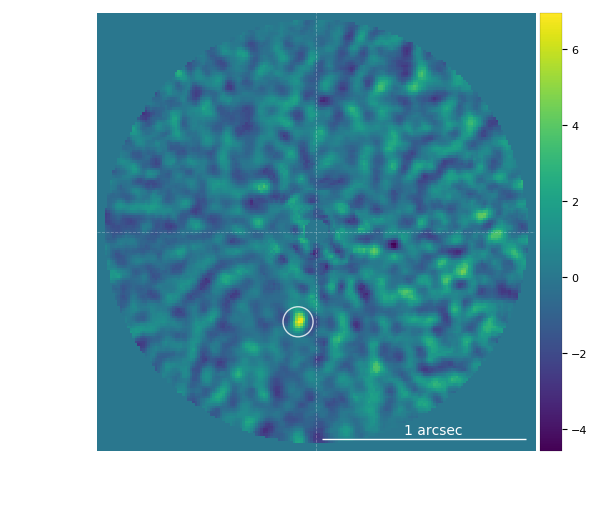}}\\
  \subfloat[LLSG S/N map]{\includegraphics[width=220pt]{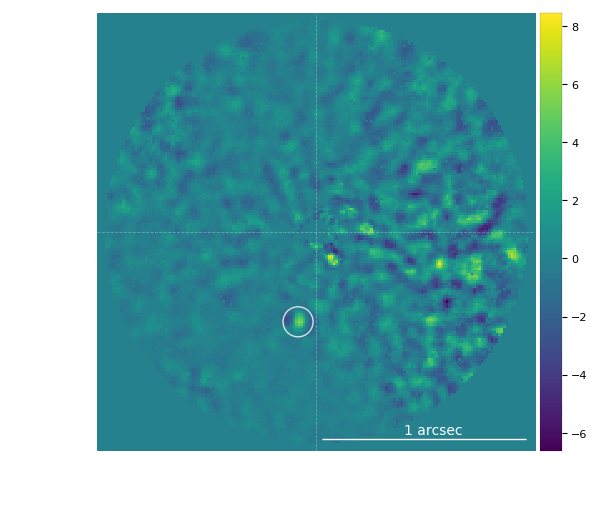}}
  \subfloat[NMF S/N map]{\includegraphics[width=220pt]{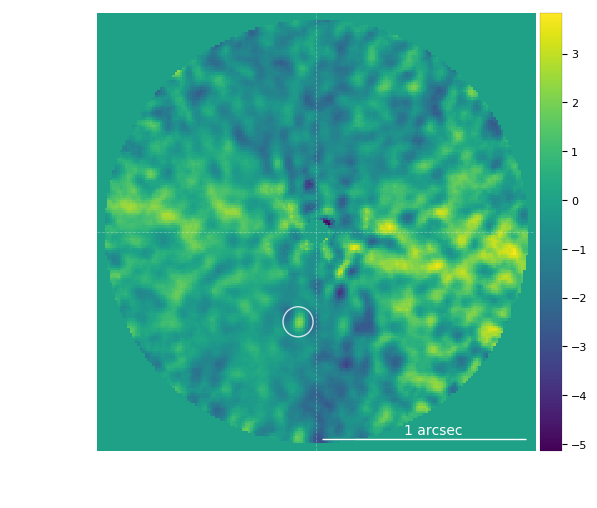}}
  \caption{\label{detmap}Probability map obtained for the SPHERE-IRDIS 51 Eridani data set, with the RSM using a Gaussian distribution along the S/N map generated with the cube of residuals obtained with annular PCA, LLSG, and NMF. The annular PCA and the NMF use 20 components,  and the LLSG has a rank of 7. The colour scale indicates the probability for the RSM map and the S/N for the three S/N maps \citep{Mawet14}. The maps are centred on the star 51 Eridani while 51 Eridani b is identified by the white circle in the lower left quadrant.}
\end{figure*}

Figure~\ref{detmap}  displays the RSM detection map and the S/N maps obtained for the SPHERE-IRDIS 51 Eridani data set (see Fig.~\ref{NACOdetmap}  for similar detection maps for the NACO $\beta$ Pictoris data set). As can be seen, the difference in intensity between the planetary signal and the background speckles is much higher with our new approach than with the usual S/N maps. 51 Eridani b \citep[contrast of $6.73 \times 10^{-6} \pm  9.02 \times 10^{-7}$ at a separation of $453.4 \pm 4.6$ mas,][]{Samland17,Maire19} can be clearly identified on the lower left quadrant with RSM, annular PCA, and LLSG, although we observe a higher number of false positives in the case of LLSG. The visual identification becomes more difficult when looking at the S/N map provided by NMF, which shows brighter wind-driven halo residuals.
                  
        \begin{figure}[t]
\begin{center}
\includegraphics[width=255pt]{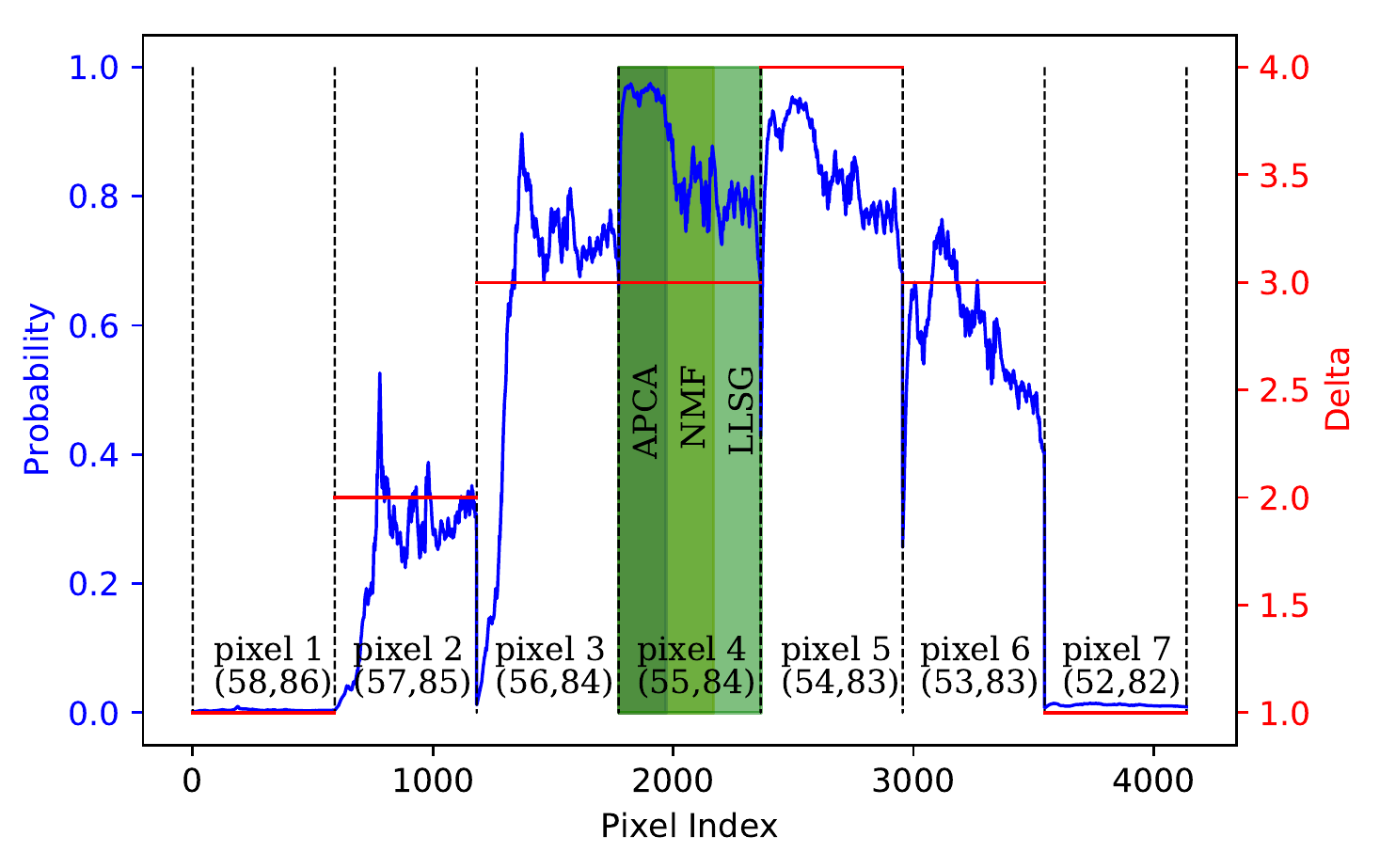} 
                        \caption{\label{probevo} Evolution of the probability in the RSM detection map around the location where 51 Eridani is detected along with the optimal $\delta$ for the respective annuli.}
                        \end{center}
                        
                          \end{figure}           

To illustrate the computation of the RSM map, Fig.~\ref{probevo} shows how the probability $\xi_{1,i_a}$  builds up when getting closer to a planetary signal; it reports the RSM map probabilities along the radial axis crossing the peak value attributed to 51 Eridani b, along with the optimal $\delta$ for the respective annuli. The data includes 7 pixels $\times$ 197 frames $\times$ 3 ADI-based post-processing techniques and is centred on the pixel showing the highest probability. As can be seen, no signal may be found in the first 591 patches representing the first pixel. The probability builds up steadily for the next three pixels until reaching a peak probability of over 95\%. The value of the optimal $\delta$ increases as well with a peak value of 4 reached at the fifth pixel, illustrating the displacement of the signal farther into the residuals distribution tail due to the increasing flux coming from the planetary candidate. We then observe a decrease of the probability and optimal $\delta$, which eventually gets back to the background speckle noise level. The stacked cube of residuals encompasses the cubes of residuals generated first by the annular PCA, then by the NMF, and finally by the LLSG. Looking at the sharp increase observed at the beginning of every pixel, we see that the strongest signal may be found in the annular PCA cube of residuals, confirming the visual analysis of the S/N maps. However the signal is still strong in the two other cubes of residuals to be able to maintain the high probability observed for the three central pixels. Changing the  order of the cubes of residuals when computing the RSM detection map does not significantly affect the probabilities.

\subsection{Receiver operating characteristic curves}

 In order to explore in more detail the properties of our new approach and compare its performance with other state-of-the-art methods, we generated synthetic data sets based on the two ADI sequences presented in the previous section. We rely on the injection of fake companions in the initial ADI sequences, an approach widely accepted by the HCI community for generating synthetic data to assess the sensitivity of post-processing methods. Since the contrast that can be reached as well as the noise structure both depend on the angular separation, we consider three different annuli as described in Table \ref{fluxval}. The comparison with the other methods is based on receiver operating characteristic (ROC) curves, which are widely used to assess the performance of binary classifiers. In these curves, one axis provides the true positive rate and the other the false positive rate. When using ROC curves for performance assessment, the main proxy for the classifier performance is the area under the ROC curve: the better the classifier, the higher the area under the ROC curve, i.e. the higher the true positive rate for a given false positive rate. We replace the false positive rate by the number of false positives for the entire frame, averaged over the number of test data sets considered for a given separation as is done in \cite{Gonzalez18}.
 
The fake companions are defined as the normalised off-axis PSF, generally measured by offsetting the target star from the coronograph, multiplied by flux values from a predefined interval defined to challenge the set of tested methods. Five different flux values are tested for each separation with step size of 0.5 times the initial value. For each flux value, eight positions are tested to mitigate the impact of bright speckles or local minima. The resulting 40 test data sets are then used to estimate the ROC curves for each separation. The contrasts for the three selected separations are provided for the NACO and SPHERE data sets in Table \ref{fluxval}. Before injecting the fake companion, we removed the known companions and some bright disc structures for the $\beta$ Pictoris data set using the negative fake companion technique \citep{Lagrange10}. We consider a false positive to be a detected companion at any other location than the one selected for the fake companion injection.

\begin{table}[t]
                        \caption{Injected companions contrast range for the three considered separations.  }
                        \label{fluxval}
\centering

                        \begin{tabular}{lll}
                        \hline
                        \hline
&\textbf{NACO} ($\beta$ Pic) & \textbf{SPHERE} (51 Eri)\\                       
 Separation   & Contrast   & Contrast  \\       
 \hline
$2$ $\lambda/D$ & 3.3-8.2 $\times 10^{-4}$ & 1.0-2.6 $\times 10^{-4}$\\
$4$ $\lambda/D$ & 0.5-1.3 $\times 10^{-4}$ & 1.2-3.1 $\times 10^{-5}$\\
$8$ $\lambda/D$  & 1.3-3.3 $\times 10^{-5}$ & 2.1-5.2  $\times 10^{-6}$\\
\hline
                        \end{tabular}
                                \end{table}

The exoplanet detections for the annular PCA, the NMF, and the LLSG methods are based on S/N maps generated using the procedure of \citet{Mawet14}.The detection of a true or false positive is done on the de-rotated median-combined individual  frame by estimating the S/N for every pixel. This estimation is done annulus-wise in order to take into account the evolution of the residuals distribution. The S/N is calculated by comparing the flux inside an aperture with a diameter of one FWHM centred on the considered pixel (i.e., 5 pixels for both data sets) with the flux of all the other apertures included in the annulus \citep[for more details about the estimation see][]{Mawet14}. Once the S/N map is computed, successive thresholds are applied onto the S/N map to create the ROC curves. For each threshold, the detection of the fake companion as well as the number of false positives are recorded and averaged over the entire set of synthetic data sets generated for the considered annulus to construct our false and true positive rates.  We follow a similar procedure for the RSM detection map,  simply replacing the S/N thresholds by successive percentage thresholds applied to the detection map.

The parameters of the different post-processing techniques have been selected to maximise the area under the ROC curves, that is, to maximise the true positive rate while minimising the number of false positives. For annular PCA and NMF, the number of principal components used to construct the reference PSF was set to 20 for both data sets. As for LLSG, we selected a rank value of 5 for the estimation of the matrix $\bm{S}$ for the $\beta$ NACO data set and 7 for the SPHERE-IRDIS data set. As regards the RSM, the mean and variance of the residuals distribution are again estimated annulus-wise. As the fake companions injected into our simulations have relatively low flux values, we tested $\delta$ in the interval $\left[ 1,5 \right]$ and kept the one leading to the highest total log-likelihood to generate the final RSM map.

\begin{figure*}
  \centering
  \subfloat[RSM Probability map]{\includegraphics[width=220pt]{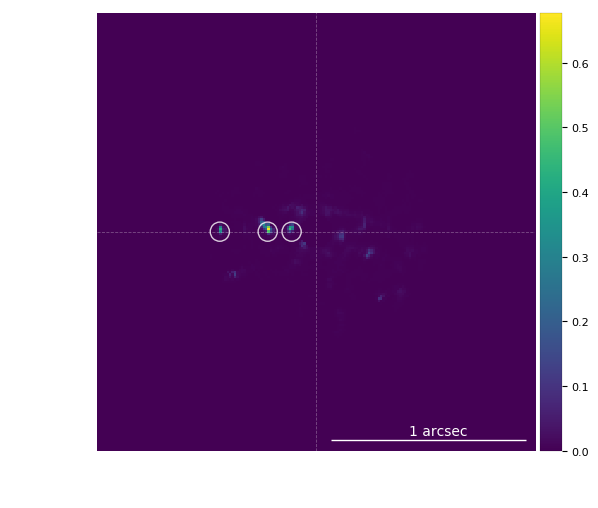}}
  \subfloat[Annular PCA S/N map]{\includegraphics[width=220pt]{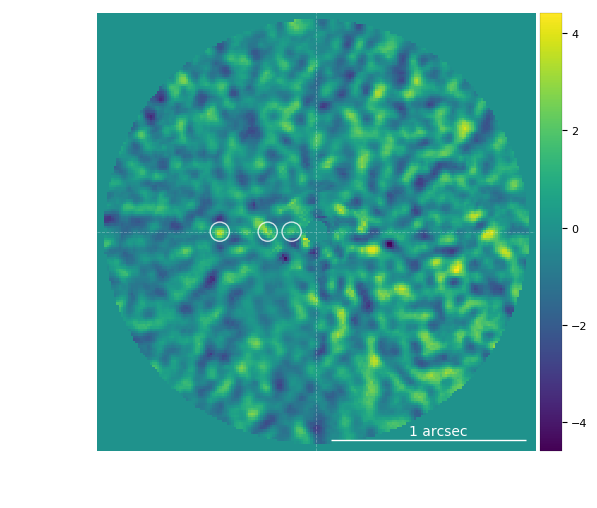}}\\
  \subfloat[LLSG S/N map]{\includegraphics[width=220pt]{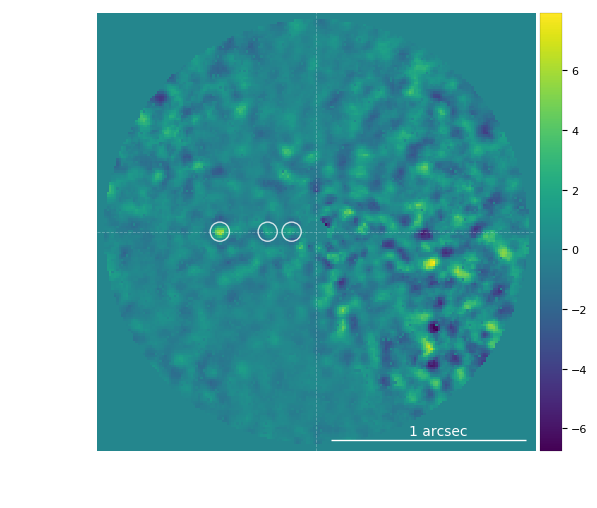}}
  \subfloat[NMF S/N map]{\includegraphics[width=220pt]{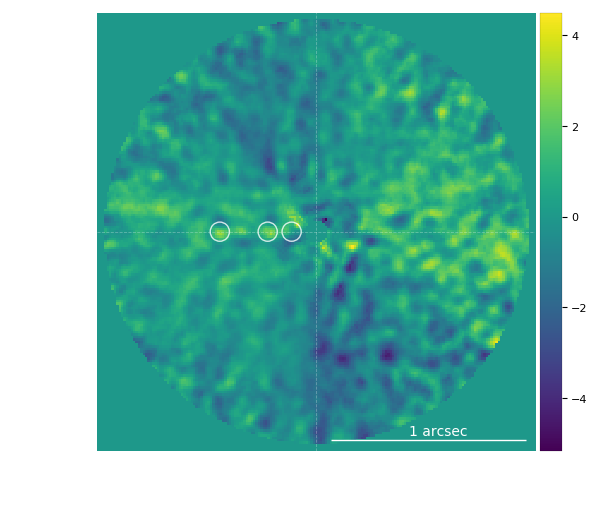}}
  \caption{\label{detmapROC} Detection map obtained after injecting three fake companions in the SPHERE-IRDIS 51 Eridani reference cube used for the ROC estimation, at a distance of 2, 4, and 8 $\lambda/D$  with a contrast of 1.0 $\times 10^{-4}$, 1.2 $\times 10^{-5}$ and 3.7  $\times 10^{-6}$,  respectively. The colour scale indicates the probability for the RSM map and the S/N for the three S/N maps. The maps are centred on the star 51 Eridani while the fake companions are identified by the white circles.}
\end{figure*}

As an illustration of the detection map calculation for the  generation of ROC curves, Fig.~\ref{detmapROC} shows the probability and S/N maps obtained by injecting fake companions with high contrast values at three different separations from the star 51 Eridani (2, 4 and 8 $\lambda/D$). As can be seen, apart from the signal injected at 8 $\lambda/D$ which appears relatively clearly in the S/N for all three post-processing methods, the RSM map is the only map providing a clear detection for all three fake companions. A set of detection maps is shown in Fig.~\ref{NACOdetmapROC} for the NACO $\beta$ Pictoris data set, leading to similar conclusions.

\subsubsection{Influence of the probability distribution}

\begin{figure*}
  \centering
  \subfloat[$\beta$ Pictoris at $2$ $\lambda/D$]{\includegraphics[width=200pt]{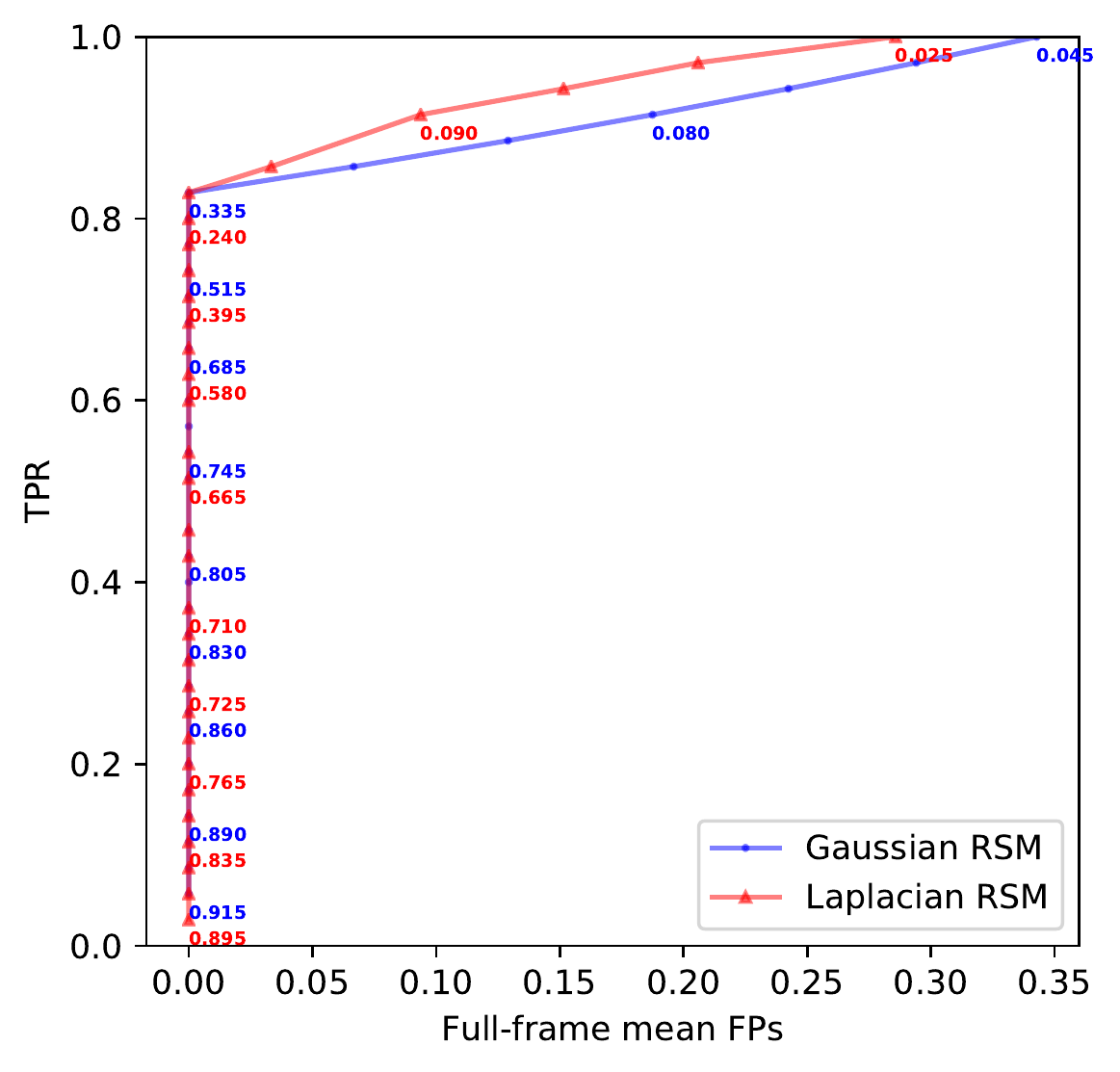}}
  \subfloat[51 Eridani at $2$ $\lambda/D$]{\includegraphics[width=200pt]{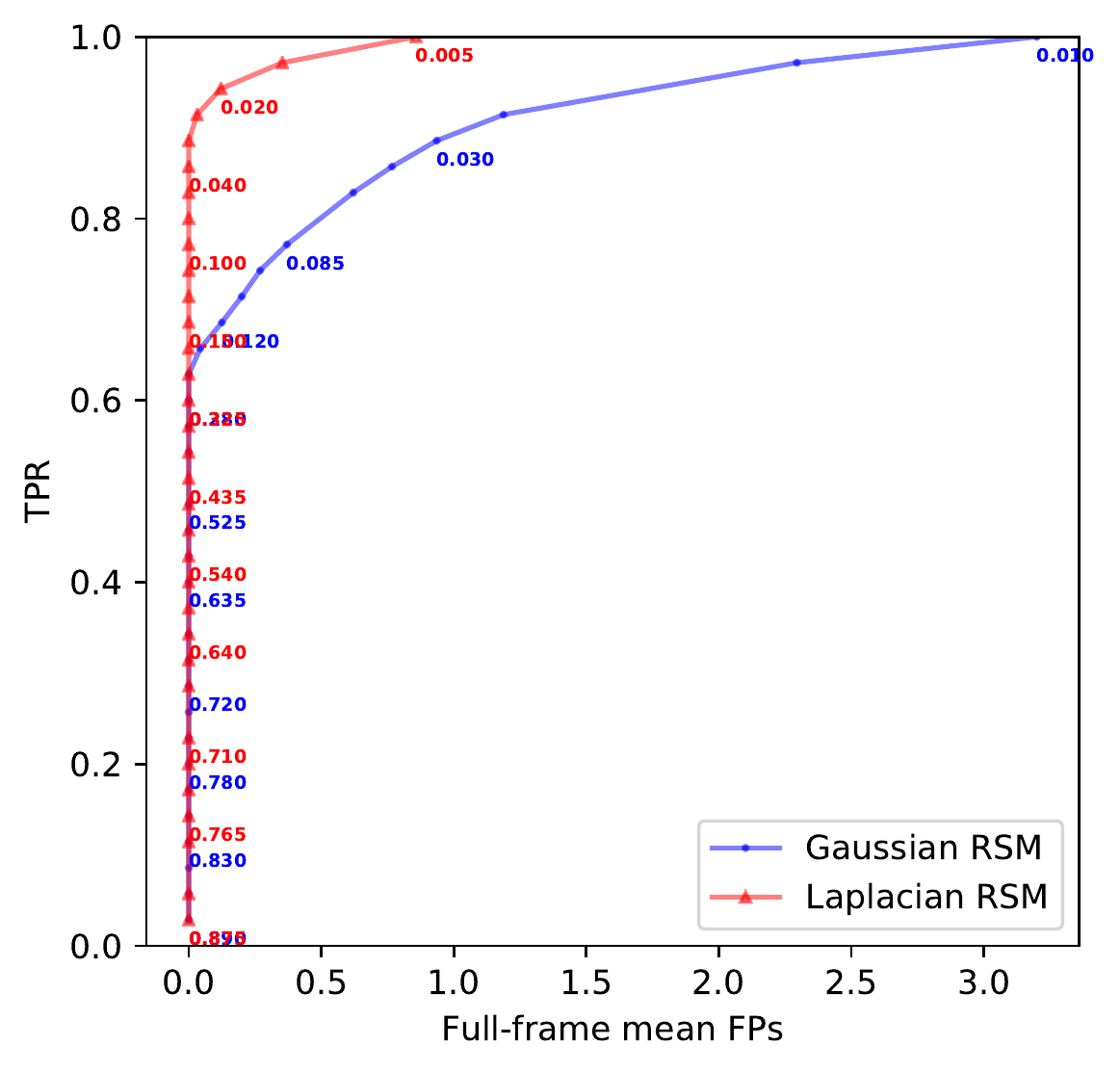}}\\
  \subfloat[$\beta$ Pictoris  at $4$ $\lambda/D$]{\includegraphics[width=200pt]{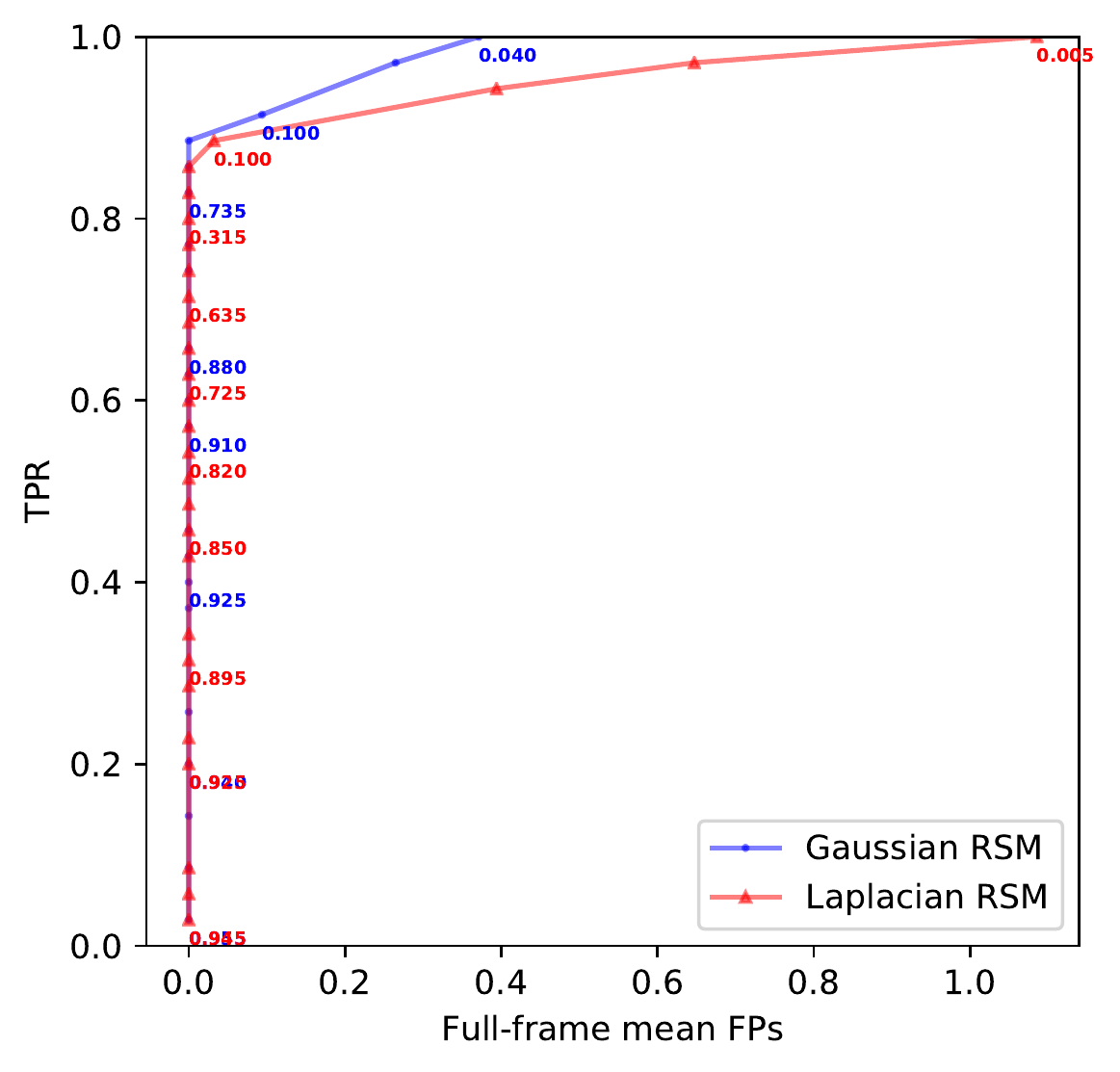}}
  \subfloat[51 Eridani at $4$ $ \lambda/D$]{\includegraphics[width=200pt]{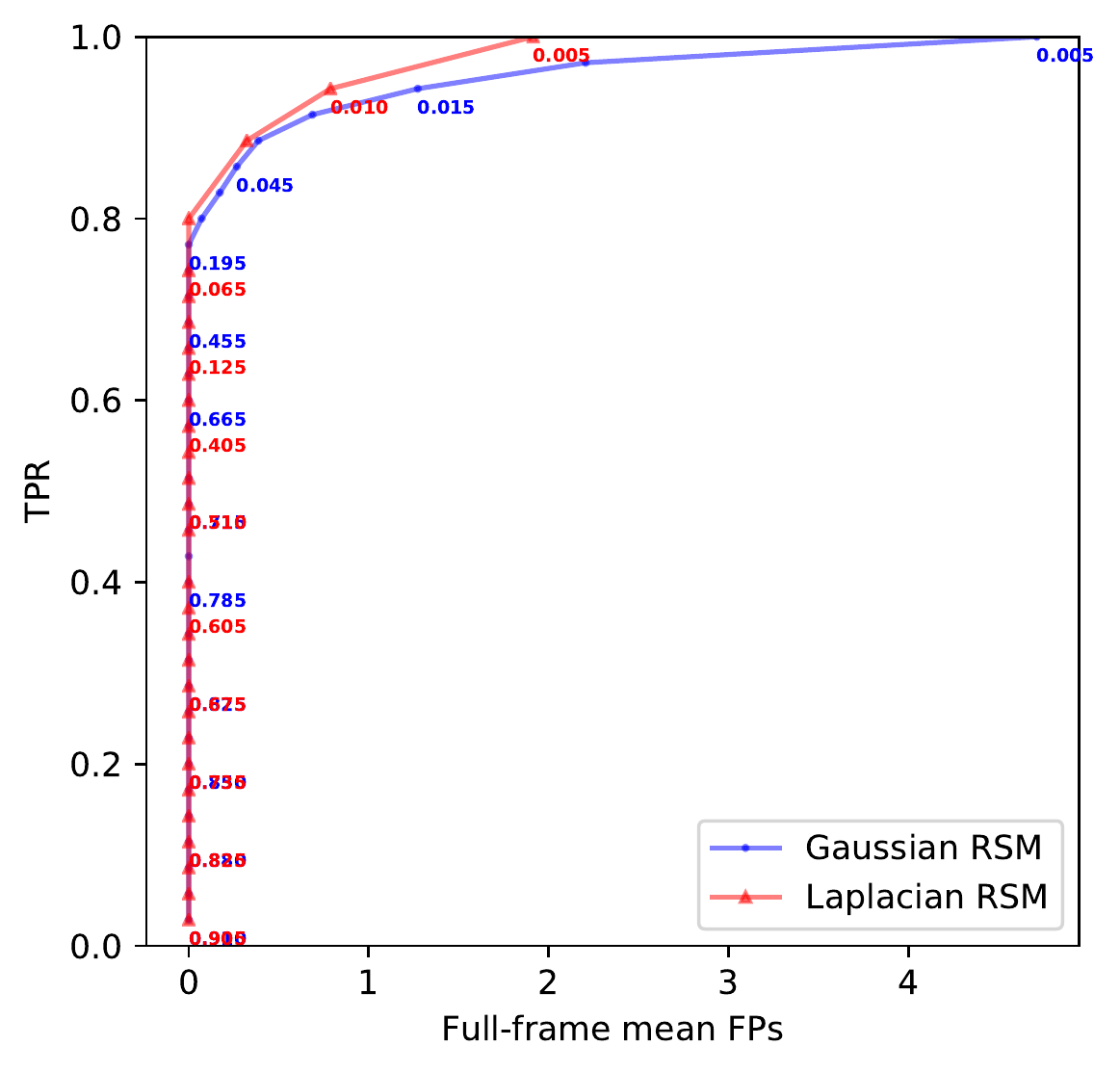}}\\
  \subfloat[$\beta$ Pictoris  at $8$ $\lambda/D$]{\includegraphics[width=200pt]{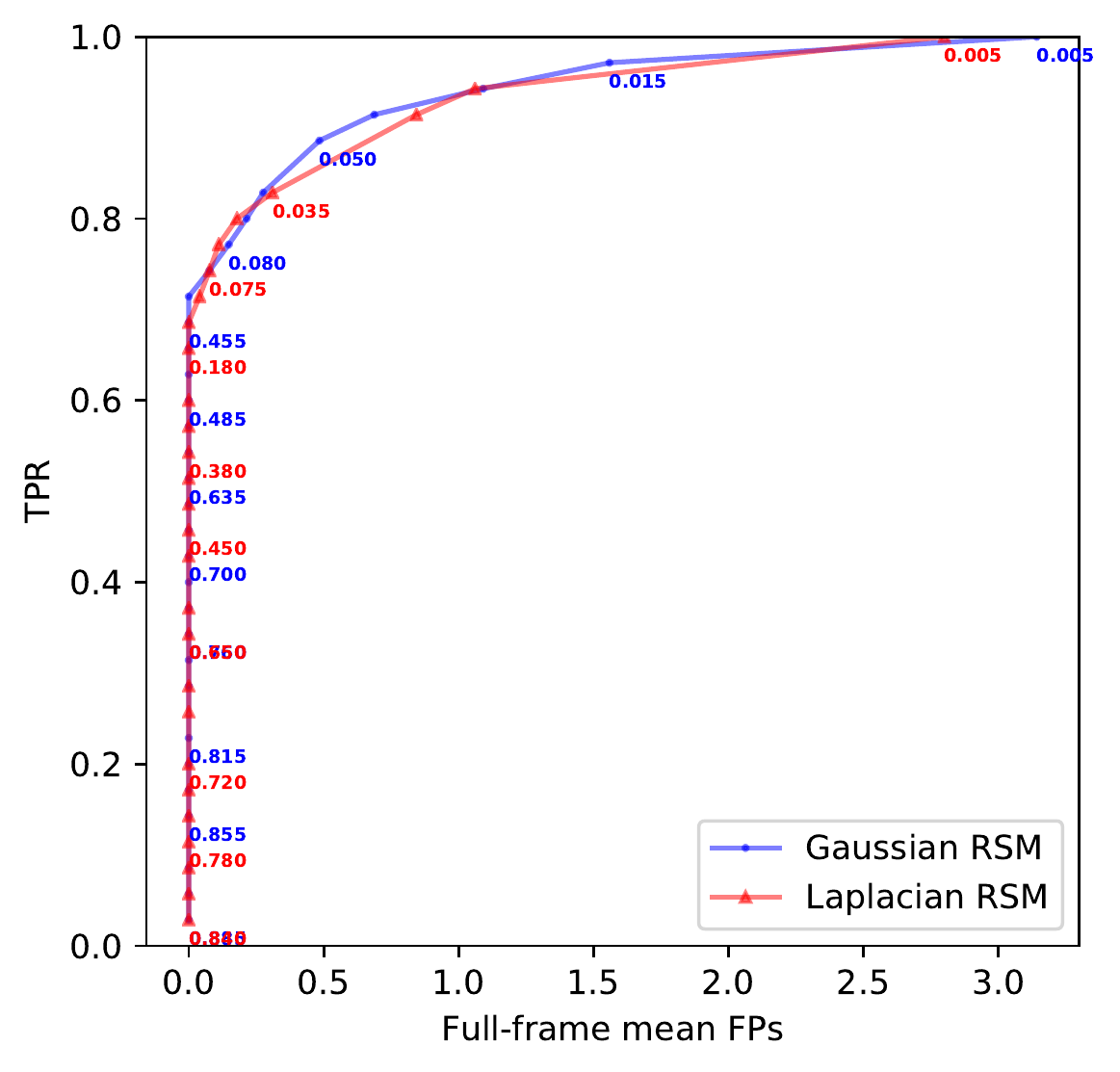}}
  \subfloat[51 Eridani at $8$ $ \lambda/D$]{\includegraphics[width=200pt]{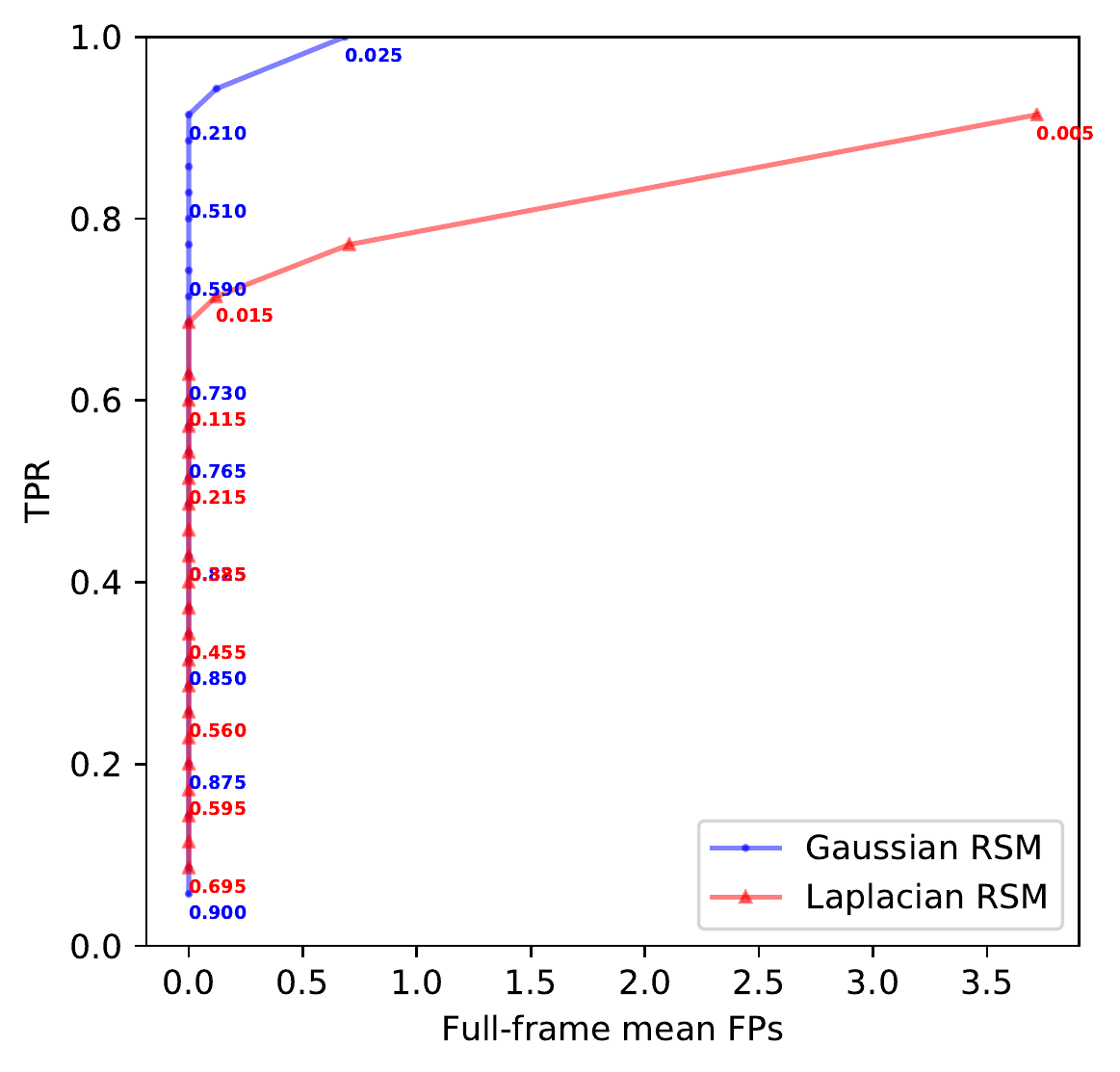}}\\ 
  
  \caption{\label{RSM} Receiver operating characteristic curves for the $\beta$ Pictoris and 51 Eridani data sets, with the RSM using a Gaussian (blue) and Laplacian (red) distribution,  respectively, to construct the likelihood function.}
\end{figure*}

We now turn to the estimation of the ROC curves which will provide more comprehensive results. We start by considering two different variants of RSM to investigate the choice of the probability distribution for the likelihood function definition. The two variants presented in  Fig.~\ref{RSM} use the Gaussian and Laplacian distribution, respectively, to construct the likelihood function appearing in $\eta_{r,i_a}$. The ROC curves are estimated for different separations; as we have seen in the previous section, the probability distribution describing the residuals evolves with angular separation. As can be seen from Fig.~\ref{RSM}, the results of the two variants are very close in the case of the $\beta$ Pictoris data set, while the distance between them becomes significant for the 51 Eridani data set. In both cases, the RSM model using the Laplacian distribution performs better for small separation while the Gaussian distribution leads to better results for larger separations. 
                          
These results confirm the findings made with Fig.~\ref{resdistri} and the importance of tails fit when selecting the optimal probability distribution. It demonstrates the interest of considering the residuals distribution evolution along the radial axis to optimally parametrise our model. We therefore propose to start the RSM detection map estimation with an analysis of the noise profile to select the right probability distribution for every separation. This additional step has been included in the RSM detection map \emph{python} package that we developed based on the model presented in this paper\footnote{The RSM detection map \emph{python} package is available on GitHub: https://github.com/chdahlqvist/RSMmap}. The function allows the user (i) to select one of the two distributions, (ii) to automatically select the best distribution based on a best-fit approach, or (iii) to create a hybrid distribution consisting in a weighted sum of both distributions. This last possibility can be useful when facing asymmetrical probability distributions as the parameters of both distributions may be estimated separately based on a best-fit approach.

\subsubsection{Comparison with S/N-based detection}
                          
\begin{figure*}
  \centering
  \subfloat[$\beta$ Pictoris at $2$ $\lambda/D$]{\includegraphics[width=200pt]{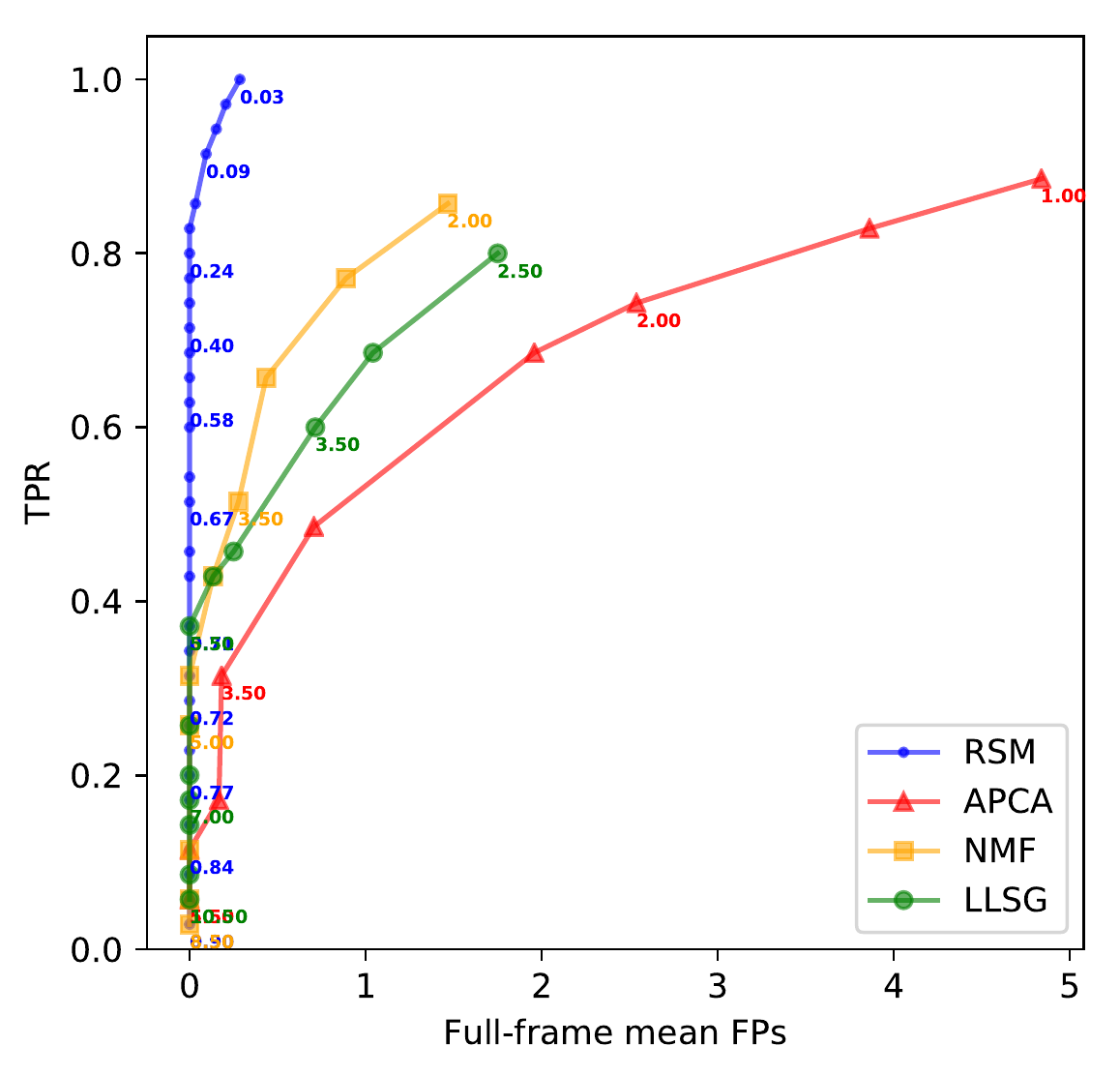}}
  \subfloat[51 Eridani at $2$ $\lambda/D$]{\includegraphics[width=200pt]{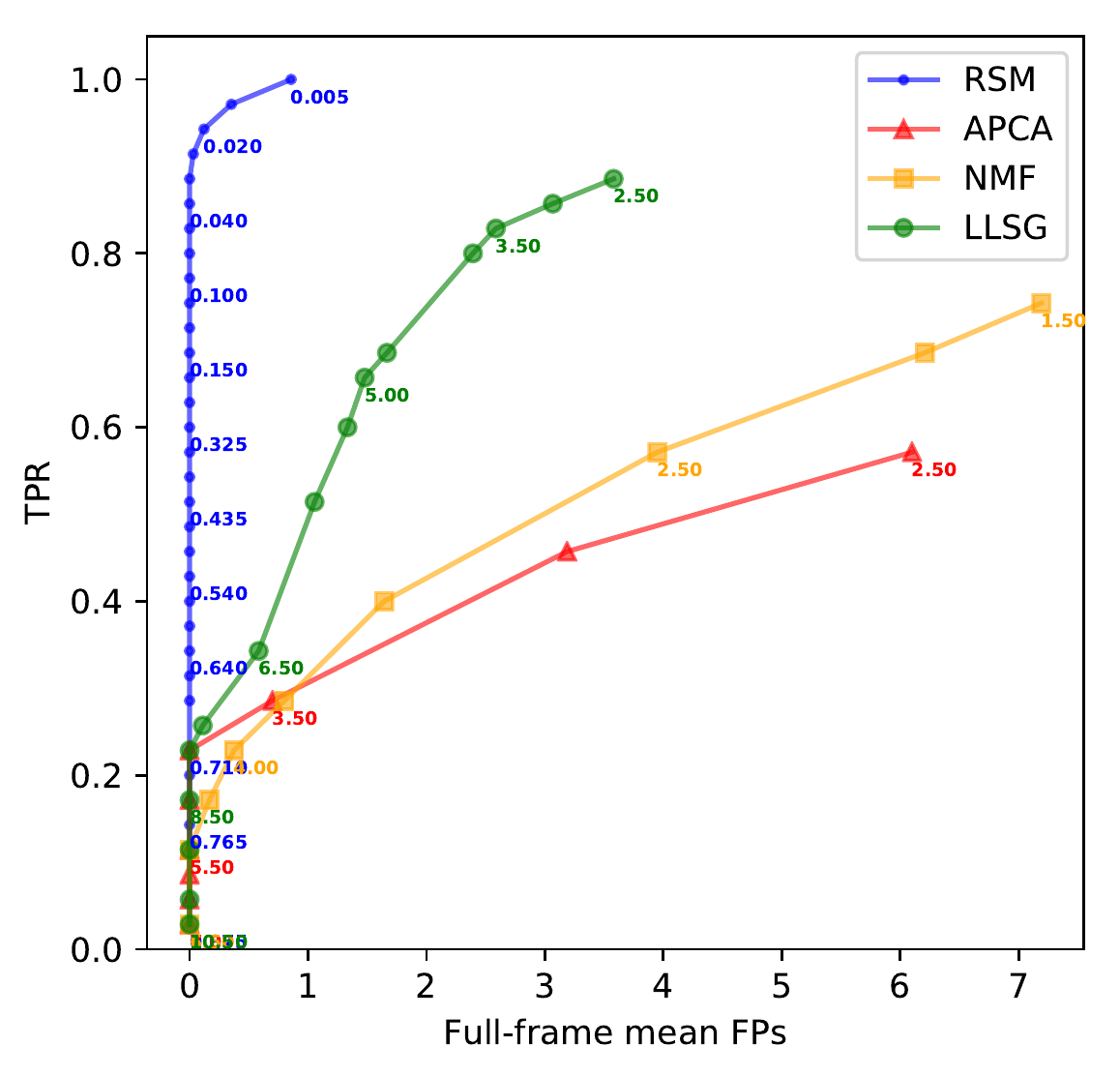}}\\
  \subfloat[$\beta$ Pictoris  at $4$ $\lambda/D$]{\includegraphics[width=200pt]{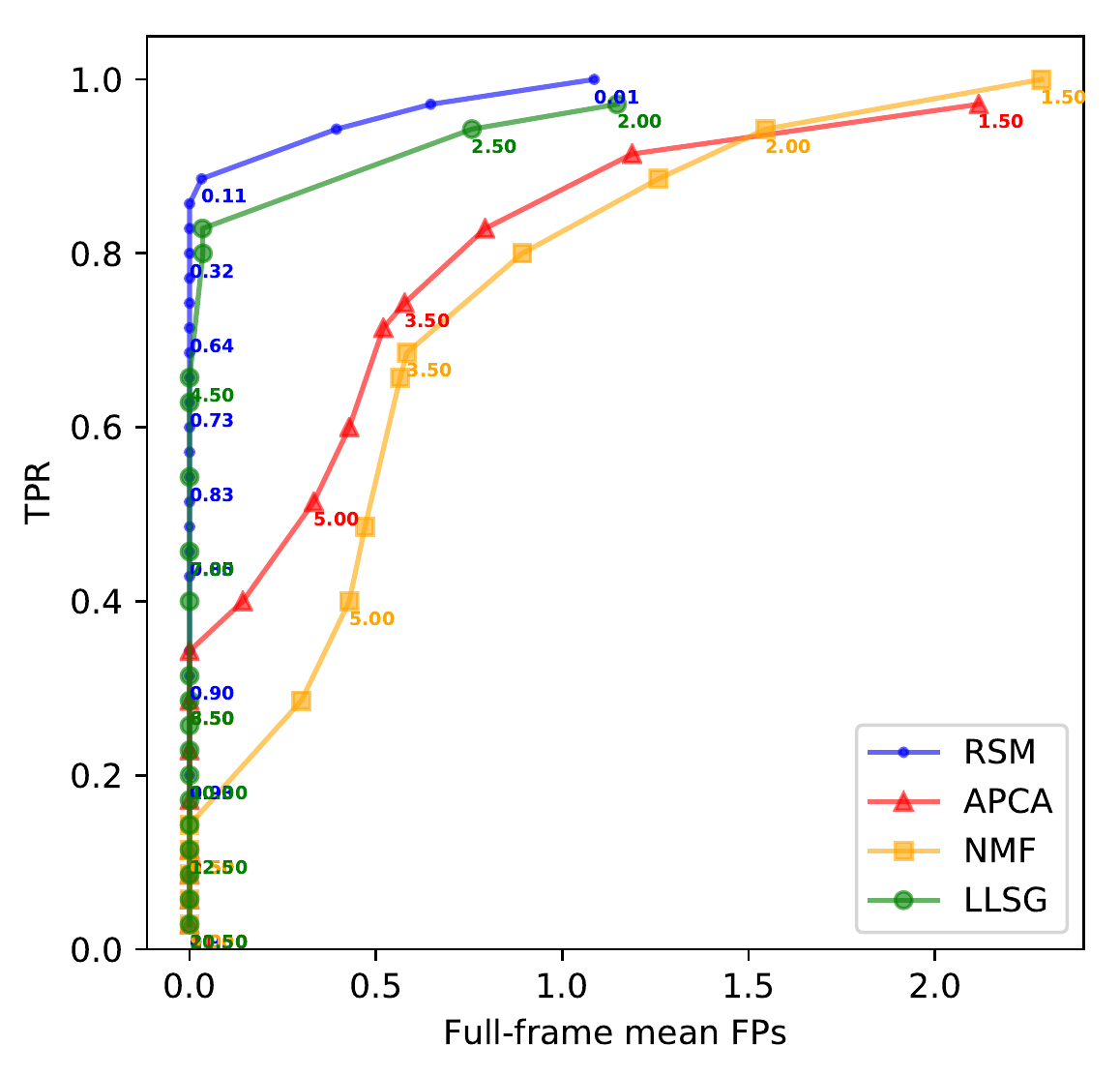}}
  \subfloat[51 Eridani at $4$ $ \lambda/D$]{\includegraphics[width=200pt]{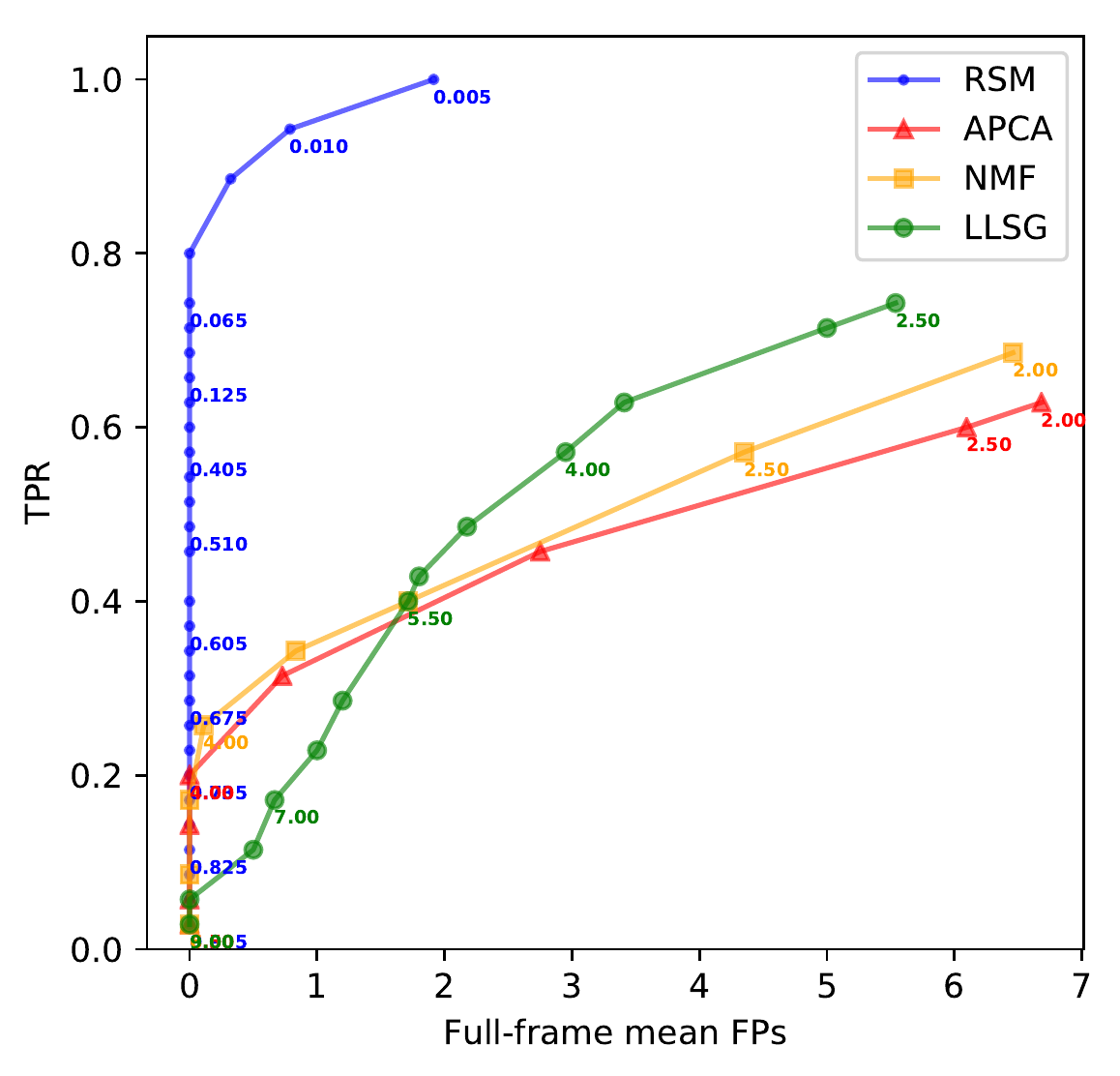}}\\
  \subfloat[$\beta$ Pictoris  at $8$ $\lambda/D$]{\includegraphics[width=200pt]{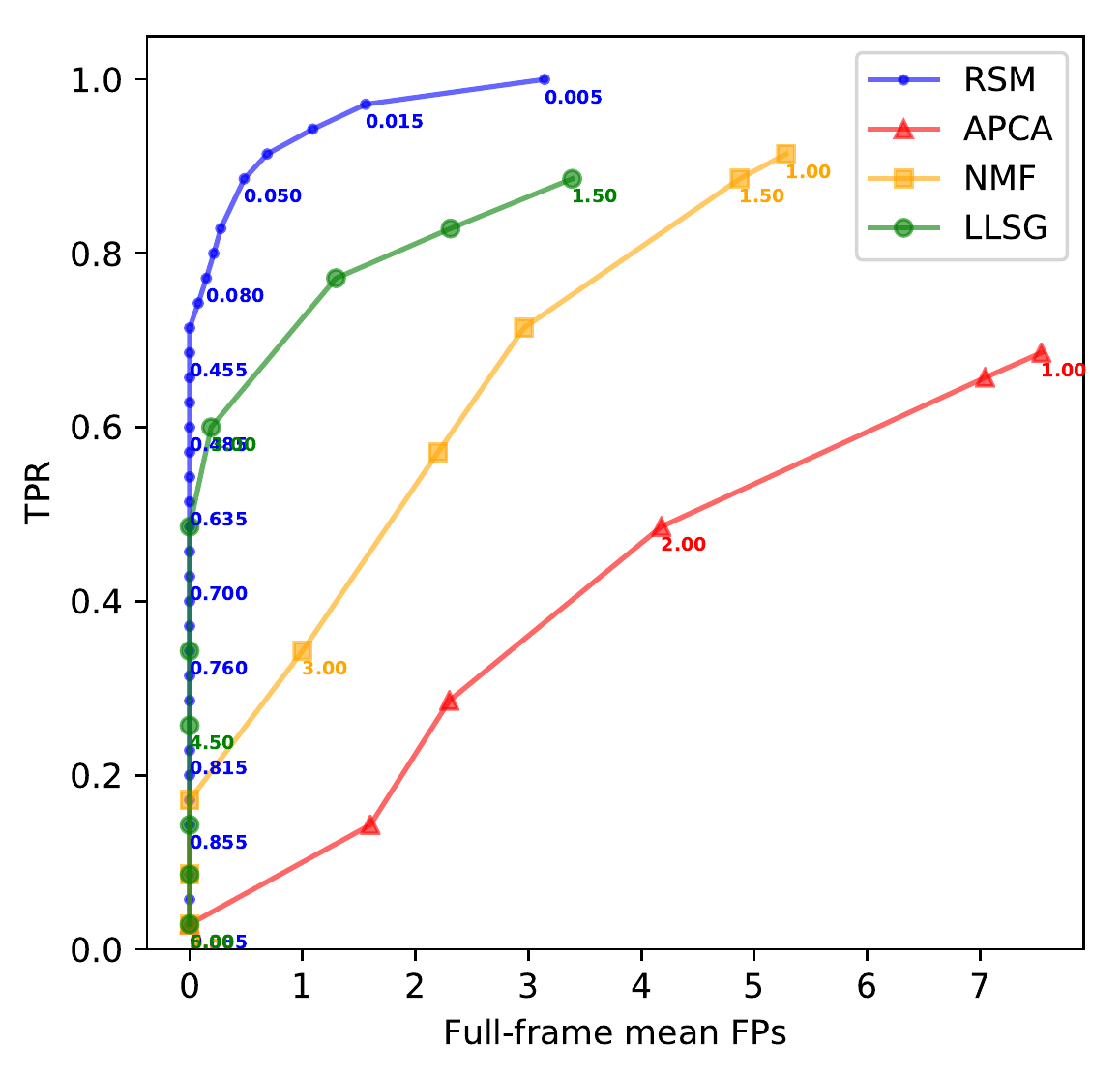}}
  \subfloat[51 Eridani at $8$ $ \lambda/D$]{\includegraphics[width=200pt]{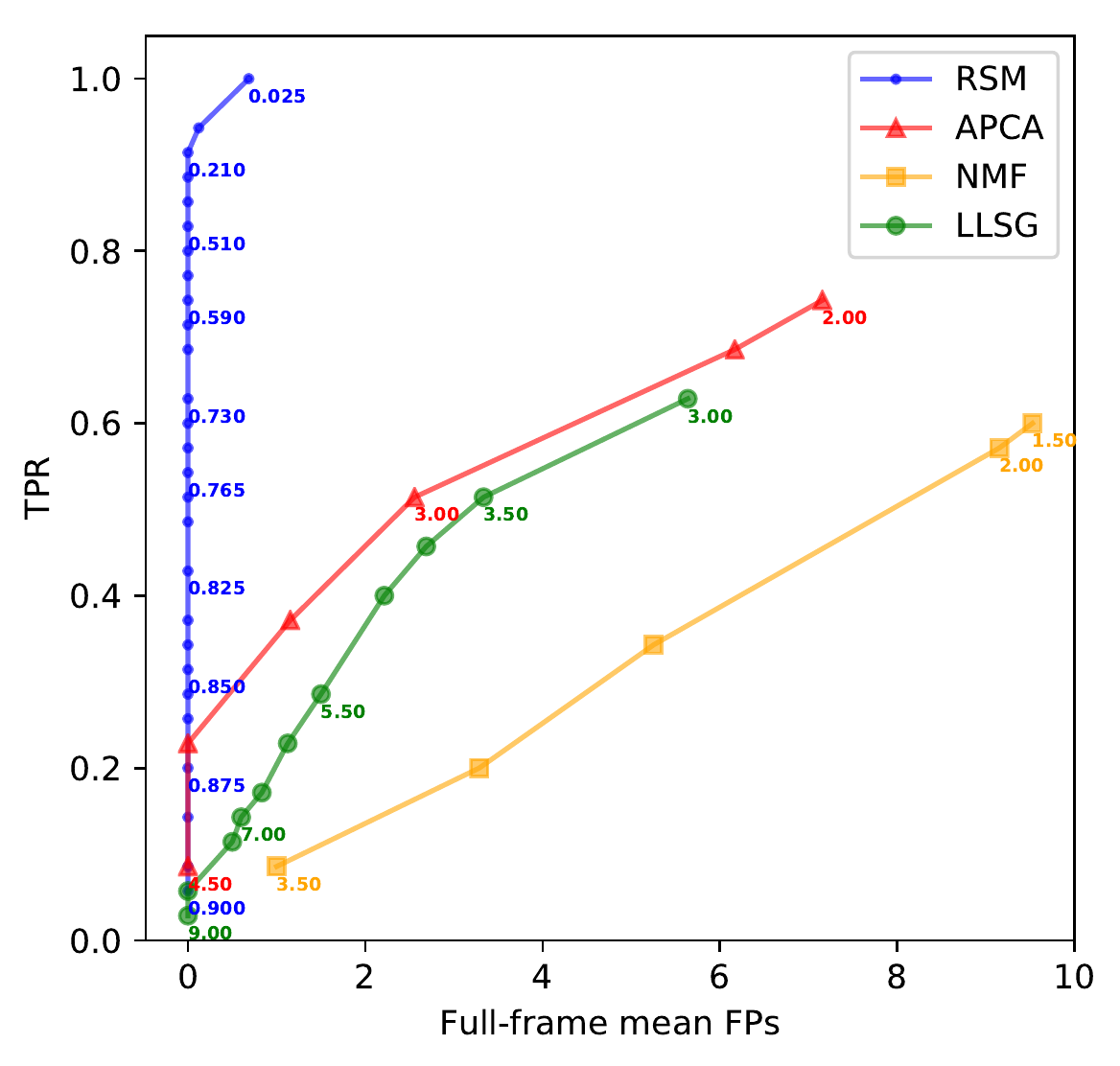}}\\ 
  
  \caption{\label{ROC}ROC curves for the $\beta$ Pictoris and 51 Eridani data sets. The RSM variant (in blue)  providing the highest area under the ROC curve in Fig.\ref{RSM} has been selected. The red, yellow, and green ROC curves are computed using the S/N map generated with respectively the annular PCA, the NMF and the LLSG.}
\end{figure*}

 We now address the question of the performance of our algorithm compared to the three post-processing methods using S/N maps. For the two data sets, Fig.~\ref{ROC} reports the ROC curves of all four methods for the same separations as before. Considering the results presented in Fig.~\ref{RSM}, we selected for each data set and each separation the distribution that provided the highest area under the ROC curve. The results demonstrate the interest of the new approach considering that the RSM performs better in every case. This may be explained by the ability of our model to be fed with multiple cubes of residuals, but also by its ability to focus only on relevant data thanks to the regime-switching feature. This allows our model to take advantage of the strength of the different post-processing methods used to produce the cubes of residuals. As speckles are not treated equally by these post-processing techniques, it is easier to remove them by taking into account several cubes of residuals. This ability to remove speckles is further improved by the memory of the RSM. Indeed, the dependence of $\xi_{s,i_a}$ on the transition matrix $p_{q,s}$ and on the probabilities at step $i_a-1$ (see Eq.~\ref{estxsi}) partly mitigates the effect of speckles on the detection map. Outliers caused by quasi-static speckles do not lead to a clear regime switch, while when facing a planetary signal the detection probability builds up along the time axis as we see in Fig.~\ref{probevo}. The dependence on the past observation reduces the noise in the final detection map significantly.

Furthermore, the possibility of selecting the right probability distribution to describe the residuals allows us to more precisely describe the behaviour of these residual speckles, which is not possible with the S/N approach. The more significant improvements for the 51 Eridani data set may be explained by the lower level of noise inside this ADI sequence, which suggests that our model should perform better with the latest generation of instruments.

\section{Conclusion}
\label{sec:conclusion}

Here, we explore the possibility of improving exoplanet detection using an RSM derived from the field of econometrics, with one regime representing the planetary signal in addition to the speckle noise and the other only the speckle noise. This novel approach allows the creation of probability maps based on cubes of residuals obtained with different ADI-based post-processing techniques. The RSM algorithm can be associated with any ADI-based post-processing techniques as it can be fed with different cubes of residuals separately or jointly. The short memory process at the heart of our RSM detection map allows quasi-static speckles to be treated more effectively when using several cubes of residuals provided by different post-processing algorithms and thereby allows the user to reach better detection performance. 

The RSM is easy to use as most of the parameters are estimated empirically. The only parameter that may need to be tuned is $\delta$, which defines the strength of the signal coming from the planetary candidates. The model automatically selects this parameter via a maximum log-likelihood approach. However, an upper value has to be defined for the interval. The estimation of the RSM map takes between three and ten times longer than the standard \citet{Mawet14} S/N map computation time, depending on the size of the ADI sequence and on the upper value for the parameter $\delta$.

We demonstrate the interest of our approach by injecting fake companions into two data sets provided by the VLT/NACO and VLT/SPHERE instruments. We compared the proposed RSM map with standard S/N maps obtained with three state-of-the-art methods: annular PCA, NMF, and LLSG. The ROC curves
clearly demonstrate the interest of our model as it outperforms all the other methods for the three angular separations considered, and for both data sets. The results also confirm that the probability distribution of the residuals evolves with radial distance and that it should be taken into account in our model when defining the likelihood function used to estimate the probability of being in one of the two regimes.  Indeed, the Laplacian distribution clearly  performs better for close separations while the Gaussian one provides better results for larger angular distances. The possibility of optimally selecting the probability distribution based on the residual noise profile has been included in the RSM detection map \emph{python} package that we have developed.

\begin{acknowledgements}
This work was supported by the Fonds de la Recherche Scientifique - FNRS under Grant n$^{\circ}$ F.4504.18 and by the European Research Council (ERC) under the European Union's Horizon 2020 research and innovation program (grant agreement n$^{\circ}$ 819155). We thank our colleague A.-L. Maire for sharing the SPHERE data set.
\end{acknowledgements}

\bibliographystyle{aa}
\bibliography{EDRSM.bib} 

\begin{appendix}

\section{NACO $\beta$ Pictoris}
\begin{figure*}
  \centering
  \subfloat[RSM Probability map]{\includegraphics[width=220pt]{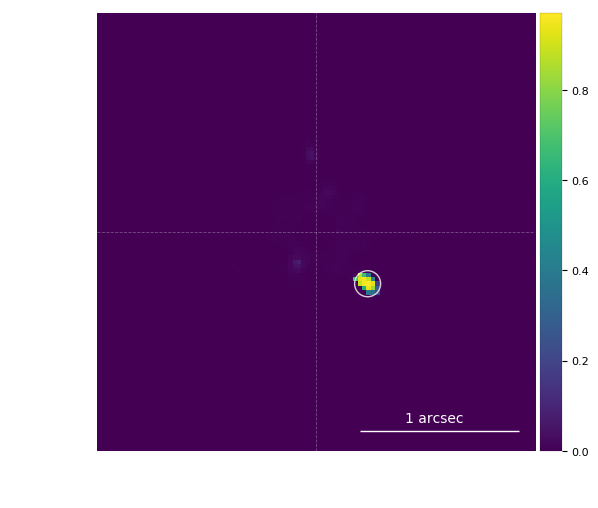}}
  \subfloat[Annular PCA S/N map]{\includegraphics[width=220pt]{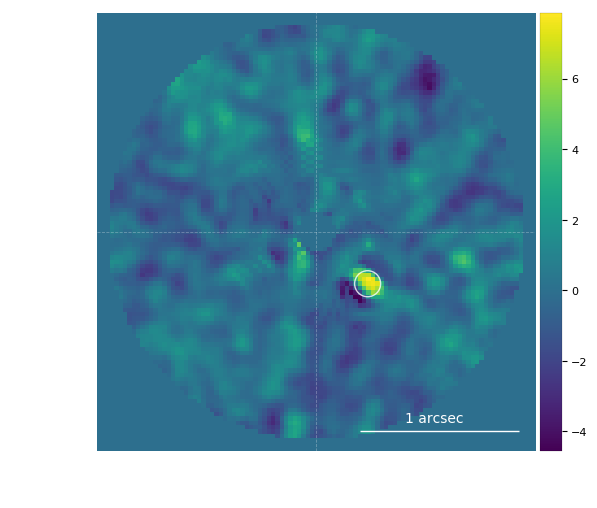}}\\
  \subfloat[LLSG S/N map]{\includegraphics[width=220pt]{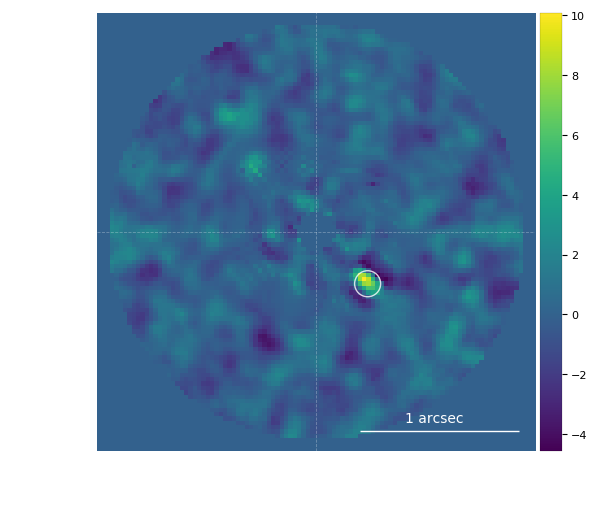}}
  \subfloat[NMF S/N map]{\includegraphics[width=220pt]{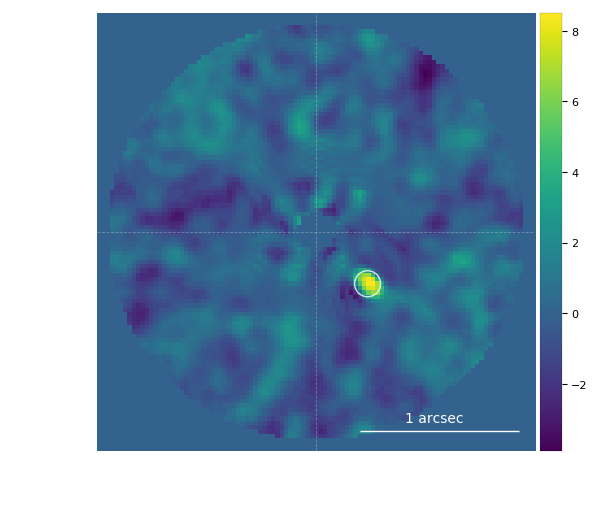}}
  \caption{\label{NACOdetmap}Probability map obtained for the NACO $\beta$ Pictoris data set, with the RSM using a Gaussian distribution along the S/N map generated with the cube of residuals obtained with annular PCA, LLSG, and NMF.  The annular PCA and the NMF use 20 components and the LLSG has a rank of 5. The colour scale indicates the probability for the RSM map and the S/N for the three S/N maps. The maps are centred on the star $\beta$ Pictoris while $\beta$ Pictoris b is identified by the white circle in the lower left quadrant.}
\end{figure*}
\begin{figure*}
  \centering
  \subfloat[RSM Probability map]{\includegraphics[width=220pt]{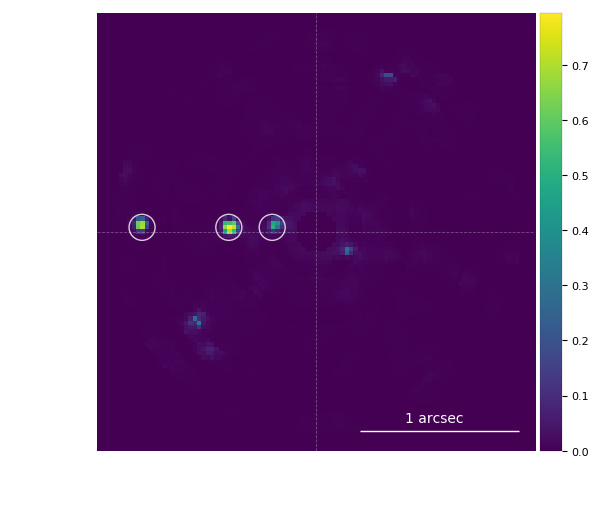}}
  \subfloat[Annular PCA S/N map]{\includegraphics[width=220pt]{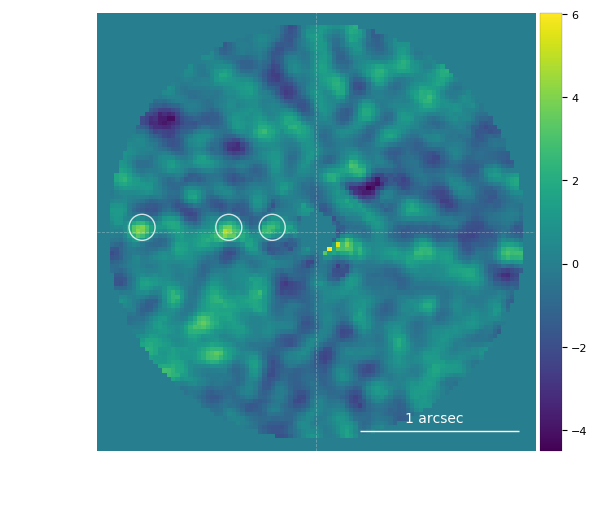}}\\
  \subfloat[LLSG S/N map]{\includegraphics[width=220pt]{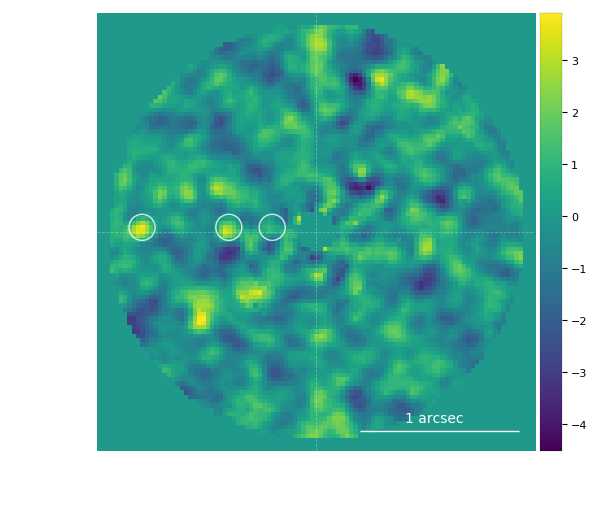}}
  \subfloat[NMF S/N map]{\includegraphics[width=220pt]{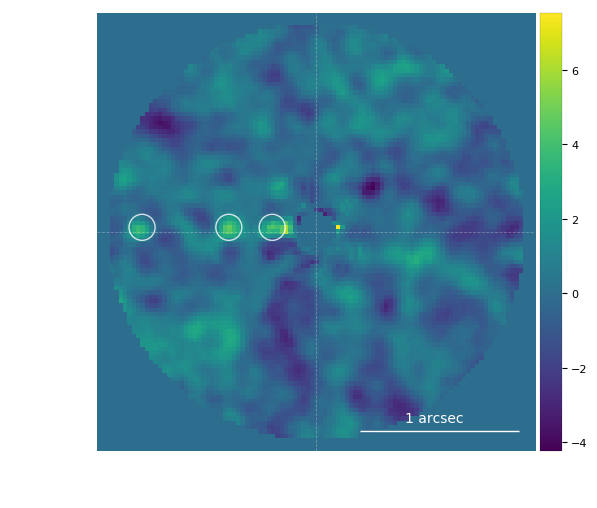}}
  \caption{\label{NACOdetmapROC}Detection map obtained after injecting three fake companions in the NACO $\beta$ Pictoris reference cube used for the ROC estimation at a distance of 2, 4, and 8 $\lambda/D$  with a contrast of 3.3 $\times 10^{-4}$, 0.4 $\times 10^{-4}$ and 1.7  $\times 10^{-5}$,  respectively. The colour scale indicates the probability for the RSM map and the S/N for the three S/N maps. The maps are centred on the star $\beta$ Pictoris while the fake companions are identified by the white circles.}
\end{figure*}

\end{appendix}

\end{document}